\documentstyle[epsfig]{mn2e}
\begin{document}
\input BoxedEPS.tex
\newcommand{\aip}{{\small ${\cal AIPS}$}}
\newcommand{\gtsim}{\mbox{{\raisebox{-0.4ex}{$\stackrel{>}{{\scriptstyle\sim}}
$}}}}
\newcommand{\ltsim}{\mbox{{\raisebox{-0.4ex}{$\stackrel{<}{{\scriptstyle\sim}}
$}}}}
\newcommand{\s}{$\stackrel{\rm s}{.}$}
\newcommand{\h}{$^{\rm h}$}
\newcommand{\m}{$^{\rm m}$}
\newcommand{\pp}{$\stackrel{\prime\prime}{.}$}
\newcommand{\de}{$^{\circ}$}
\newcommand{\p}{$^{\prime}$}
\newcommand{\arc}{$^{\prime\prime}$}
\newcommand{\marc}{^{\prime\prime}}
\newcommand{\rs}{{\em $r_s$}}
\newcommand{\DPM}{{\em DPM}}
\newcommand{\alf}{{\displaystyle\biggl({\nu_{\rm h} \over \nu_{\rm l}}\biggr)^{\alpha}} }

\newcommand{\figstart}[1]
    { \begin{figure}[htb]
      \begin{picture}(0,#1) }
\newcommand{\figend}[4]
    { \end{picture}
      \special{#1}
      \caption[#2]{#3}
      \label{#4}
      \end{figure} }
\newcommand{\fig}[5]
    { \figstart{#1}
      \figend{#2}{#3}{#4}{#5} }
\newcommand{\bHS}{\beta_{\mbox{\scriptsize HS}}}
\newcommand{\bBF}{\beta_{\mbox{\scriptsize BF}}}
\newcommand{\nT}{\nu_{\mbox{\scriptsize T}}}
\newcommand{\et}{E_{\mbox{\scriptsize T}}}
\newcommand{\nTn}{\nu_{\mbox{\scriptsize Tn}}}
\newcommand{\nTf}{\nu_{\mbox{\scriptsize Tf}}}
\newcommand{\tn}{\tau_{x\mbox{\scriptsize n}}}
\newcommand{\tf}{\tau_{x\mbox{\scriptsize f}}}
\newcommand{\xn}{x_{\mbox{\scriptsize n}}}
\newcommand{\xf}{x_{\mbox{\scriptsize f}}}
\newcommand{\yn}{y_{\mbox{\scriptsize n}}}
\newcommand{\yf}{y_{\mbox{\scriptsize f}}}
\newcommand{\lln}{l_{\mbox{\scriptsize n}}}
\newcommand{\llf}{l_{\mbox{\scriptsize f}}}
\newcommand{\Dn}{f(\Delta_{\mbox{\scriptsize n}})}
\newcommand{\Df}{f(\Delta_{\mbox{\scriptsize f}})}
\newcommand{\B}{\mbox{$B$}}
\newcommand{\Bo}{\mbox{$B$}_{0}}

\SetEPSFDirectory{/scratch/sbgs/figures/hst/}
\SetRokickiEPSFSpecial
\HideDisplacementBoxes

\title[Revised SWIRE photometric redshifts]{Revised SWIRE photometric redshifts}
\author[Rowan-Robinson M. et al]{Michael Rowan-Robinson$^1$, Eduardo Gonzalez-Solares$^2$, Mattia Vaccari$^{3,4}$,
\newauthor
Lucia Marchetti$^3$\\
$^1$Astrophysics Group, Blackett Laboratory, Imperial College of Science 
Technology and Medicine, Prince Consort Road,\\ 
London SW7 2AZ,\\
$^2$Institute of Astronomy, Madingley Rd, Cambridge CB3 0HA,\\
$^3$Dipartimento di Fisica e Astronomia  "Galileo Galilei", Universita du Padova, Vicolo Osservatorio 3, I-35122 Padua, Italy,\\
$^4$Department of Physics, University of Western Cape, 7535 Belliville, Cape Town, South  Africa,\\
}
\maketitle
\begin{abstract}
We have revised the SWIRE Photometric Redshift Catalogue to take account of new optical photometry
in several of the SWIRE areas, and incorporating 2MASS and UKIDSS near infrared data.  Aperture
matching is an important issue for combining near infrared and optical data, and we have explored
a number of methods of doing this.  The increased number of photometric bands available for the
redshift solution results in improvements both in the rms error and, especially, in the outlier rate.

We have also found that incorporating the dust torus emission into the QSO templates improves the
performance for QSO redshift estimation.  Our revised redshift catalogue contains over 1 million
extragalactic objects, of which 26288 are QSOs.

\end{abstract}


\section{Introduction}

Rowan-Robinson et al (2008) reported photometric redshifts for over 1 million galaxies in the
Spitzer SWIRE survey (the SWIRE Photometric Redshift Catalogue, hereafter SPRC) and gave a detailed review of earlier work on photometric redshifts.
Subsequently new results have been reported by Wolf et al (2008), Brammer et al (2008), 
Ilbert et al (2009), Salvato et al (2009), and a comparison of photometric redshift methods 
has been published by by Hildebrandt et al (2010).

The SWIRE Survey consisted of 49 sq deg of sky surveyed by Spitzer at 3.6, 4.5, 5.8, 8.0, 24.0, 70.0 and 160.0 $\mu$m.
The SWIRE areas in which we had optical photometry and were able to derive photometric redshifts in SPRC were:
(1)  8.72 sq deg of ELAIS-N1 (EN1), in which we had 5-band (U'g'r'i'Z') photometry from the Wide Field Survey (WFC), (2) 4.84 sq deg of ELAIS-N2 (EN2), in which we had 5-band 
(U'g'r'i'Z') photometry from the WFS, 
(3) 7.53 sq deg of the Lockman Hole, in which we had 3-band photometry (g'r'i') from the SWIRE photometry 
programme, with U-band photometry in 1.24 sq deg, (4) 4.56 sq deg in Chandra Deep Field South 
(CDFS), in which we had 3-band (g'r'i') photometry from the SWIRE photometry programme, (5) 6.97 sq 
deg of the XMM Large Scale Structure Survey (XMM-LSS), in which we had 5-band (UgriZ) photometry from Pierre et al (2007). 
(6) 1.5 sq deg of ELAIS-S1, in which we have 3-band (B,V,R) photometry from Berta et al (2007).  
In addition within XMM 
we had 10-band photometry (ugrizUBVRI) from the VVDS programme of McCracken et al (2003), Le F\'evre et al (2005) (0.79 sq deg), 
and very deep 5-band photometry (BVRi'z') in 1.12 sq deg of the Subaru XMM Deep Survey 
(SXDS, Furusawa et al 2008).   The SWIRE data are described in Surace et al (2004) \footnote[1]{and at
http:$//$irsa.ipac.caltech.edu$/$data$/$SPITZER$/$SWIRE$/$}.  The estimated redshifts range from 0 to 6 but the bulk are at
z $<$ 1.5.

The advent of revised INT WFC UgriZ optical fluxes for Lockman, EN1 and EN2 (Gonzalez-Solares et al 2011), 
the release of CFHT Legacy Survey T0005 MegaCam ugriz optical fluxes for XMM 
(http://terapix.iap.fr/cplt/oldSite/Descart/CFHTLS-T0005-Release.pdf)
and the UKIDSS DR8 release of WFCAM JK fluxes for Lockman, EN1 and XMM-LSS (Lawrence et al. 2007 \footnote[2]{http://surveys.roe.ac.uk/wsa/dr8plus-release.html}), make it worthwhile revisiting the SWIRE photometric redshifts in these areas.

The Spitzer-selected 'data fusion'catalogues compiled by Vaccari et al. (2012 in prep) comprises most publicly available
photometric and spectroscopic data such as the above in most fields surveyed by Herschel as part of the
HerMES survey (Oliver et al. 2012).
TOPCAT \footnote[3]{http://www.star.bris.ac.uk/$\sim$mbt/topcat/} was used to merge the SPRC catalogues for Lockman,
EN1, EN2 and XMM-LSS with the data fusion catalogues, with a 1.5" search radius, to get the desired photometric  data.
Not all sources from the original SPRC found matches, mostly because the data fusion selection requires
a source to be detected at either 3.6 or 4.5 $\mu$m and some 24 $\mu$m sources from the original SPRC
are thus missing from the data fusion catalogues.  Table 1 summarizes the optical and near infrared data available
in each SWIRE area.

\begin{table*}
\caption{Explanation of magnitude fields}
\begin{tabular}{llll}
field  &  am1-5    &       am21,22,23,25,26 & am6-8\\
&&&\\
EN1   &   rev. WFC UgriZ  &     SDSS ugriz &   2MASS$/$UKIDSS JHK\\
EN2    &  rev. WFC UgriZ   &     SDSS  ugriz &  2MASS JHK\\
Lock  &  rev. SWIRE Ugri &    SDSS  ugriz &  2MASS$/$UKIDSS JHK\\
VVDS  &  VVDS ugri   &    UBVRI  & VVDS JK\\
SXDS  &  Subaru      &    Megacam ugriz & -\\
XMM   &  SWIRE  UgriZ     &    Megacam ugriz & UKIDSS JHK\\
CDFS  &  SWIRE gri      &    -    &   -\\
S1   &   SWIRE B,V,R    &    -    &   -\\
\end{tabular}
\end{table*}

Rowan-Robinson et al (2008) used optical magnitudes and Spitzer IRAC 3.6 and 4.5 $\mu$m fluxes to 
estimate photometric redshifts.  They reported difficulty in incorporating 2MASS and UKIDSS J,H,K magnitudes
into the solution,which they attributed to issues of aperture matching.  These problems have been solved here.  
The use of these additional bands, together with the improved optical photometry, has resulted in a reduction in
the number of catastrophic outliers and improved rms values, when photometric and spectroscopic redshifts 
are compared.  Since redshifts are now determined from up to 15 bands, compared with generally a maximum 
of 7 previously, the new redshifts are more reliable.  The new catalogues comprise 1,009,607 redshifts, 
out of a total of 1,066,879 in the original SPRC.

\section{Aperture corrections}
Aperture matching between wavelength bands is crucial to the success of photometric redshifts.  For distant
galaxies photometry in a 2 or 3 arcsec aperture will give the integrated light from the whole galaxy.  For
nearby galaxies photometry in the same aperture would be dominated by light from the central regions of the
galaxy and might comprise only a few $\%$ of the integrated light.  In template-fitting methods such as we are 
using here it is natural to try to seek an estimate of the integrated spectral energy distribution (SED), so that
near and distant galaxies can be fitted with the same template.  This also has the benefit that derived properties
such as luminosity, star-formation rate, stellar mass and dust mass have a physical meaning for the galaxy.

There are several options for estimating the integrated light from an extended galaxy in any particular waveband.
Optical and near infrared catalogues generally provide Kron and Petrosian integrated magnitudes.  
Sextractor (Bertin and Arnouts 1996) provides a mag-auto integrated magnitude (similar to Kron magnitude).  These 
integrated magnitudes are derived essentially by a curve of growth fitted to
photometry derived in a series of apertures of different sizes.  However in practice using integrated magnitude 
estimates for each photometric band to derive the integrated SED gives poor results for photometric redshifts.
This is presumably because of the uncertainty introduced by the process of estimating the integrated magnitude,
primarily because of different contributions of sky photon noise.
A much more successful option is to start from photometry derived in a single small aperture in each band, and
then apply an aperture correction derived in a single chosen band to all the bands.  This is the approach followed by
Rowan-Robinson et al (2008) and Gonzalez-Solares et al (2011).

\subsection{Optical data}
The optical photometry in SPRC had been derived in most areas using Sextractor and we used magnitudes
measured in a 2" diameter aperture, applying an r-band aperture correction\\

\medskip
	delmag$_{SPRC}$ =  r(mag-auto) - r,  	(1)\\

where r is measured in a 2" aperture (WFC 'aper2'),

to all bands.  All optical magnitudes used in this work are AB magnitudes and all J,H,K magnitudes are
Vega magnitudes (with the exception of the VVDS area, where they are AB magnitudes).  The quoted optical and
near infrared magnitudes have been PSF (point spread function) aperture corrected, to correct for the light lying outside 
the aperture for a point source.  Hereafter in this paper we use the term 'aperture correction'
to mean the extended source aperture correction.
Fig 1L shows the optical aperture correction, delmag$_{SPRC}$, used in 
SPRC, versus redshift for the Lockman area, essentially a plot of (inverse) angular size versus redshift.  The aperture
correction was only applied to the optical magnitudes if it lies in the range -0.10 to -5.0.   Otherwise it was set to zero.
Sources with aperture correction $>$ -0.10 mag are considered to be point sources within the uncertainties of 
the photometric accuracy and the estimation of the integrated magnitude.  Sources with aperture correction $<$ -5.0 mag. 
($<0.01\%$ of sources) are considered to be erroneous determinations of integrated magnitude.

For SDSS optical data, available for the Lockman, EN1 and EN2 areas, the 'model' magnitude provides a well-calibrated
integrated magnitude.  Fig 1R shows delmag$_{SPRC}$ versus r(SDSS,model)-r(WFC,mag-auto).  Quasars almost all have no aperture 
correction, as expected.  For galaxies there is no sign of any correlation of r(SDSS,model)-r(WFC,mag-auto) with delmag$_{SPRC}$, so the
SDSS model magnitude is estimating approximately the same total magnitude as the WFC mag-auto.
However there is quite a wide dispersion in r(SDSS,model)-r(WFC,mag-auto), sufficient to harm photometric redshift
estimation (see SED plots below, Fig 14).  This dispersion arises from the uncertainties in the two estimates of the integrated magnitude.
 We have therefore added r(WFC,mag-auto)-r(SDSS,model) to the SDSS model magnitudes
to normalize the two sets of aperture corrections.  The reason for making
the WFC r-magnitude the preferred choice is that there are far more SWIRE sources with WFC data than with
SDSS data.  Also the WFC goes one magnitude deeper than SDSS and therefore provides more accurate magnitudes for 
fainter SDSS sources.   For SDSS data we investigated using the point-source (PS) magnitudes, with aperture correction 
r(SDSS,Petr)-r(SDSS,PS), but this gave inferior results to using the SDSS model magnitudes.  This is understandable
because the model magnitudes use an aperture derived in the r-band and then measure the flux in that aperture in
all the bands, ensuring consistent galaxy colours.

\begin{figure*}
\epsfig{file=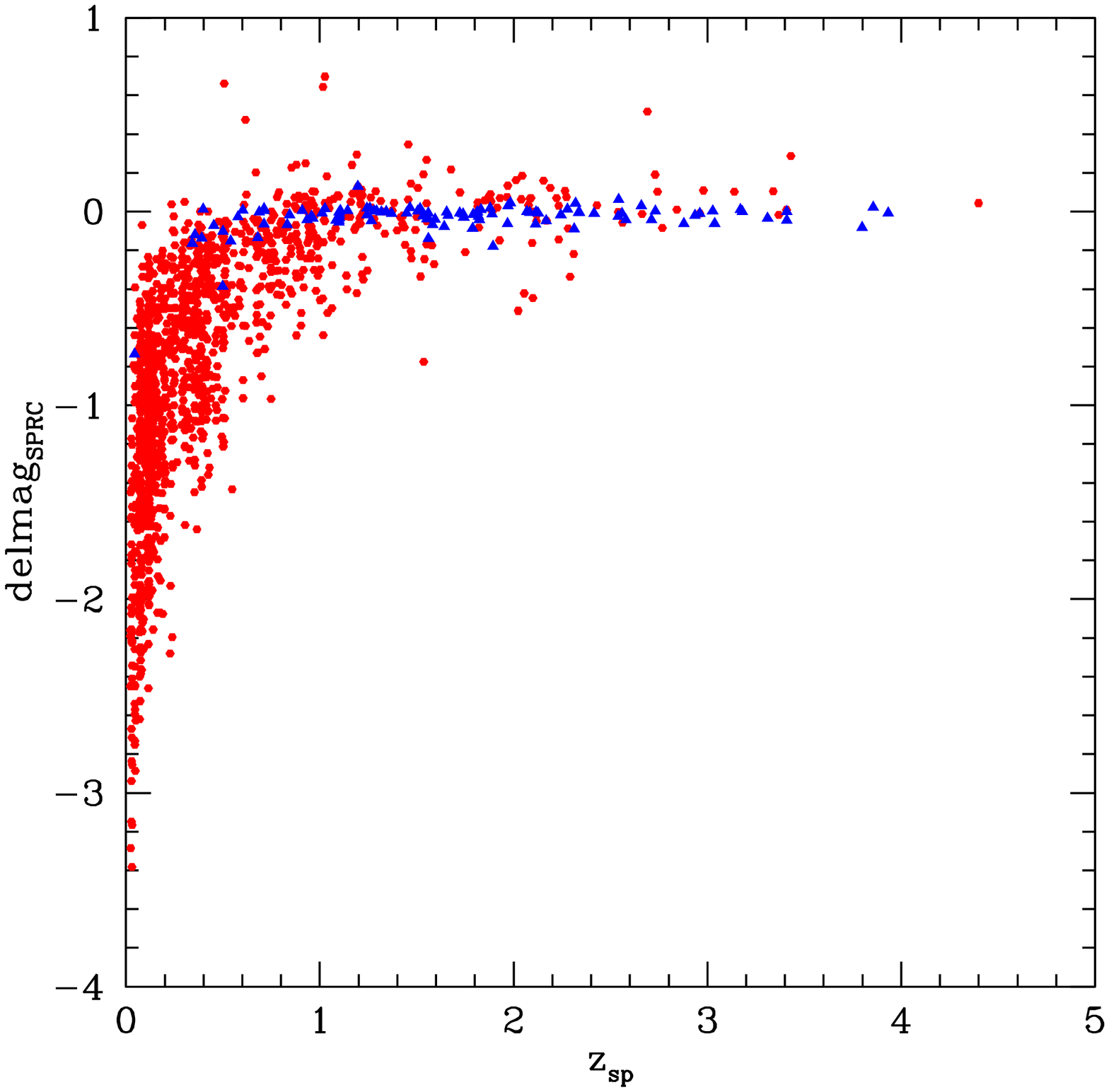,angle=0,width=7cm}
\epsfig{file=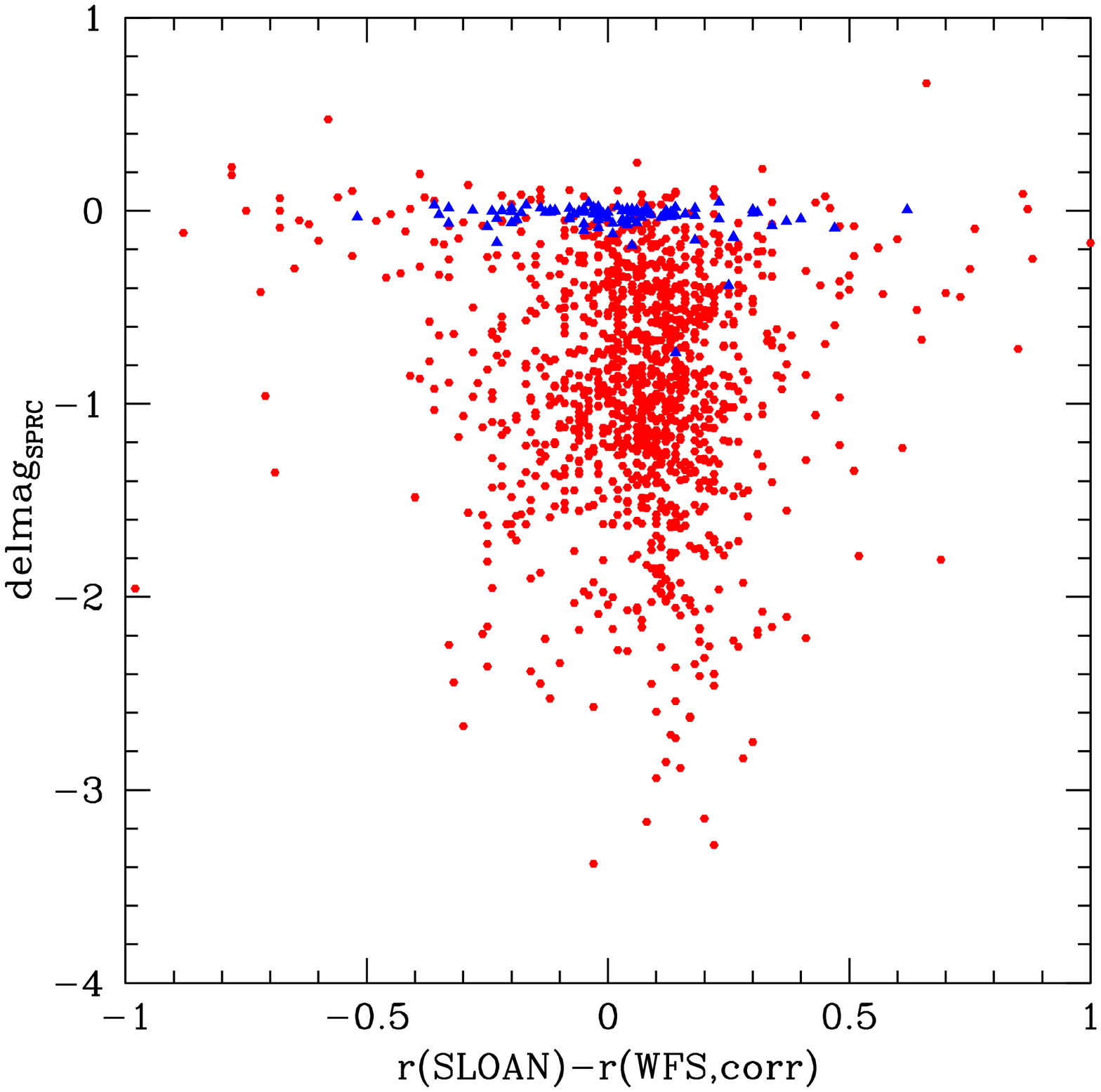,angle=0,width=7cm}
\caption{LH: SPRC aperture correction versus $z_{sp}$ for Lockman.
RH: SPRC aperture correction versus r(SDSS,model)-r(WFC,mag-auto).  Red symbols: galaxies,
blue symbols: QSOs.}
\end{figure*}

We investigated various other options for aperture corrections in the optical.  The revised WFC photometry of
Gonzalez-Solares et al (2011), available in the fusion catalogues, includes the Petrosian magnitude in
each band, so we define\\

	delmag$_{WFC}$ = r(WFC,Petr)-r.		(2)\\

This is well correlated with delmag$_{SPRC}$ (Fig 2L), but with significant scatter.

\begin{figure*}
\epsfig{file=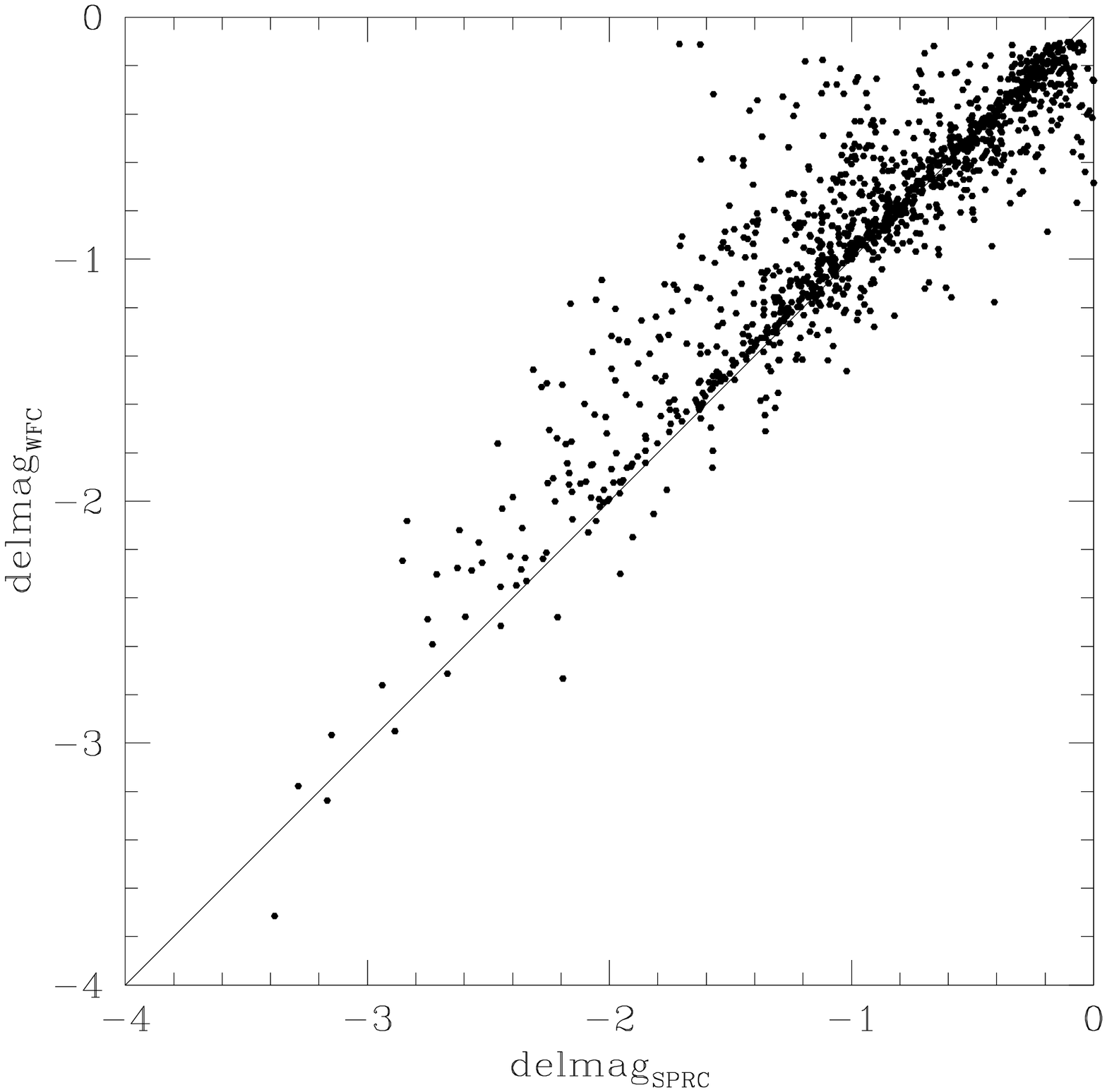,angle=0,width=7cm}
\epsfig{file=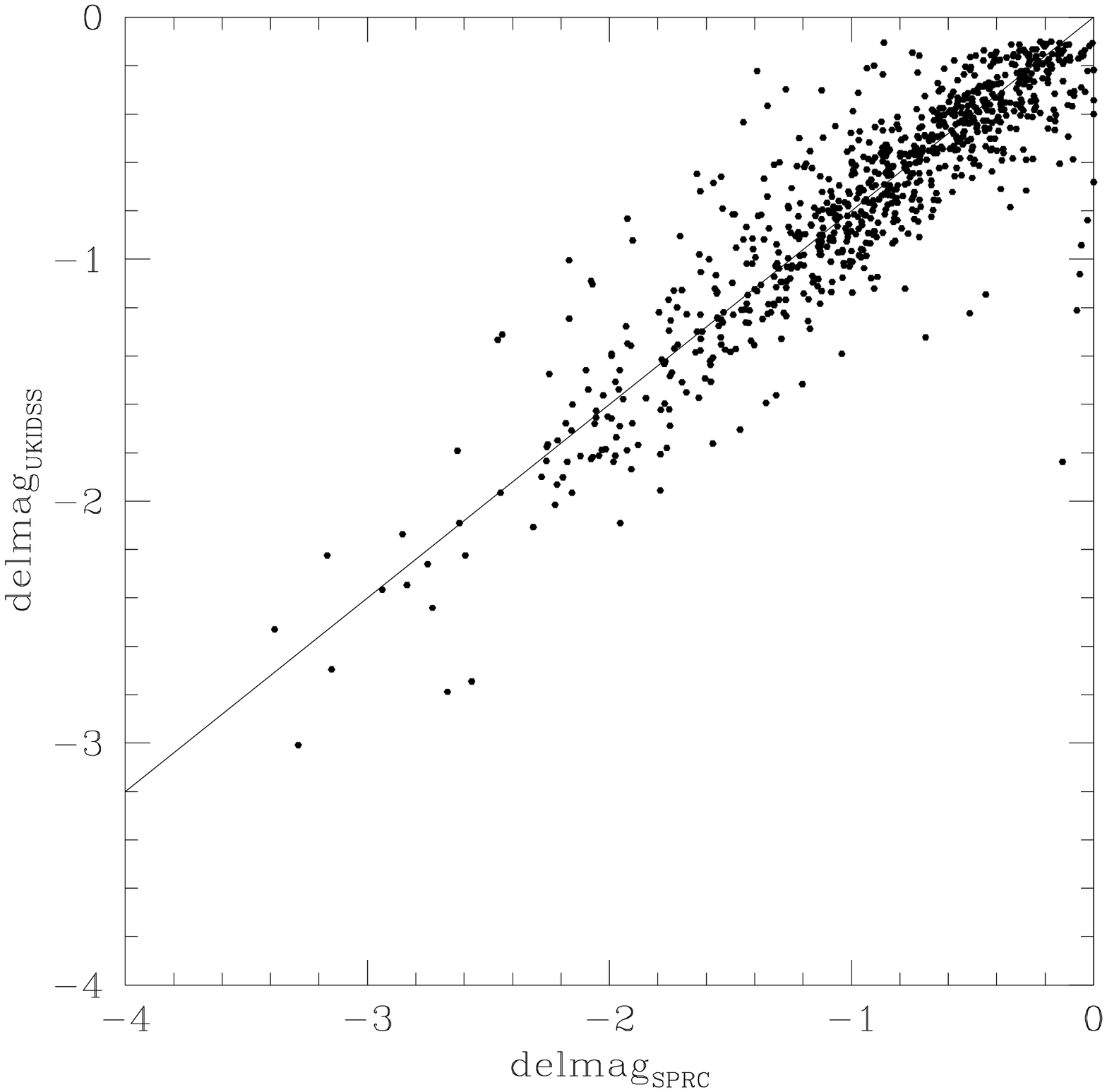,angle=0,width=7cm}
\caption{LH: Plot of delmag$_{WFC}$ (eqn 2) versus delmag$_{SPRC}$, the SPRC aperture correction, in Lockman.  The straight
line has slope 1.
RH: Plot of delmag3 (eqn 3) versus delmag$_{SPRC}$, the SPRC aperture correction. The straight line has slope 0.8.
}
\end{figure*}

Using delmag$_{WFC}$ instead of delmag$_{SPRC}$ led to slightly worse phot-z results, so we decided to stick with delmag$_{SPRC}$, which
was derived from the Sextractor r-band mag-auto, for WFC data (while using the revised WFC magnitudes
supplied in the fusion catalogue).  However where a mag-auto estimate is  not available, we have used 
delmag$_{WFC}$.

\subsection{Near infrared data}

 For 2MASS data an option for the aperture correction is to use the K-iso magnitude if available,
and the K-PS magnitude otherwise, with aperture correction\\

\medskip
	delmag$_{2MASS}$=K(2MASS,iso)-K(2MASS,PS). 	(3)\\

For UKIDSS data the natural aperture correction to consider is\\

\medskip
	delmag$_{UKIDSS}$ =K(UKIDSS,Petr)-K(UKIDSS,aper3),           (4)\\
	
applied to the aper3 (1.0") J,K magnitudes. 
	
 Both delmag$_{2MASS}$ and delmag$_{UKIDSS}$ are quite well correlated with delmag$_{SPRC}$ (Fig 2R shows the correlation of delmag$_{UKIDSS}$ with 
delmag$_{SPRC}$) and this suggests the idea of using k*delmag$_{SPRC}$ as the near infrared aperture correction, with k to be determined 
for each survey.  The direct use of delmag$_{2MASS}$ and delmag$_{UKIDSS}$ for 2MASS and UKIDSS magnitudes, respectively, 
resulted in a worse phot-z solution, so the use of k*delmag$_{SPRC}$ was explored in some detail.  

Fig 3 shows delmag$_{SPRC}$ versus (z-J), with no aperture correction applied to J.  There is a very strong correlation.
Fig 4L shows delmag$_{SPRC}$ versus (z-J) for 2MASS data, with an aperture correction 0.8*delmag$_{SPRC}$ applied to J.
Fig 4R shows  delmag$_{SPRC}$ versus (z-J) for UKIDSS data, with an aperture correction 1.1*delmag$_{SPRC}$ applied to J.
Only sources with spectroscopic redshift $<$0.3 have been included to minimize the possibility of evolution of colour
with angular size.
We can see that these corrections work well in removing the correlation of colour with delmag$_{SPRC}$.

\begin{figure}
\epsfig{file=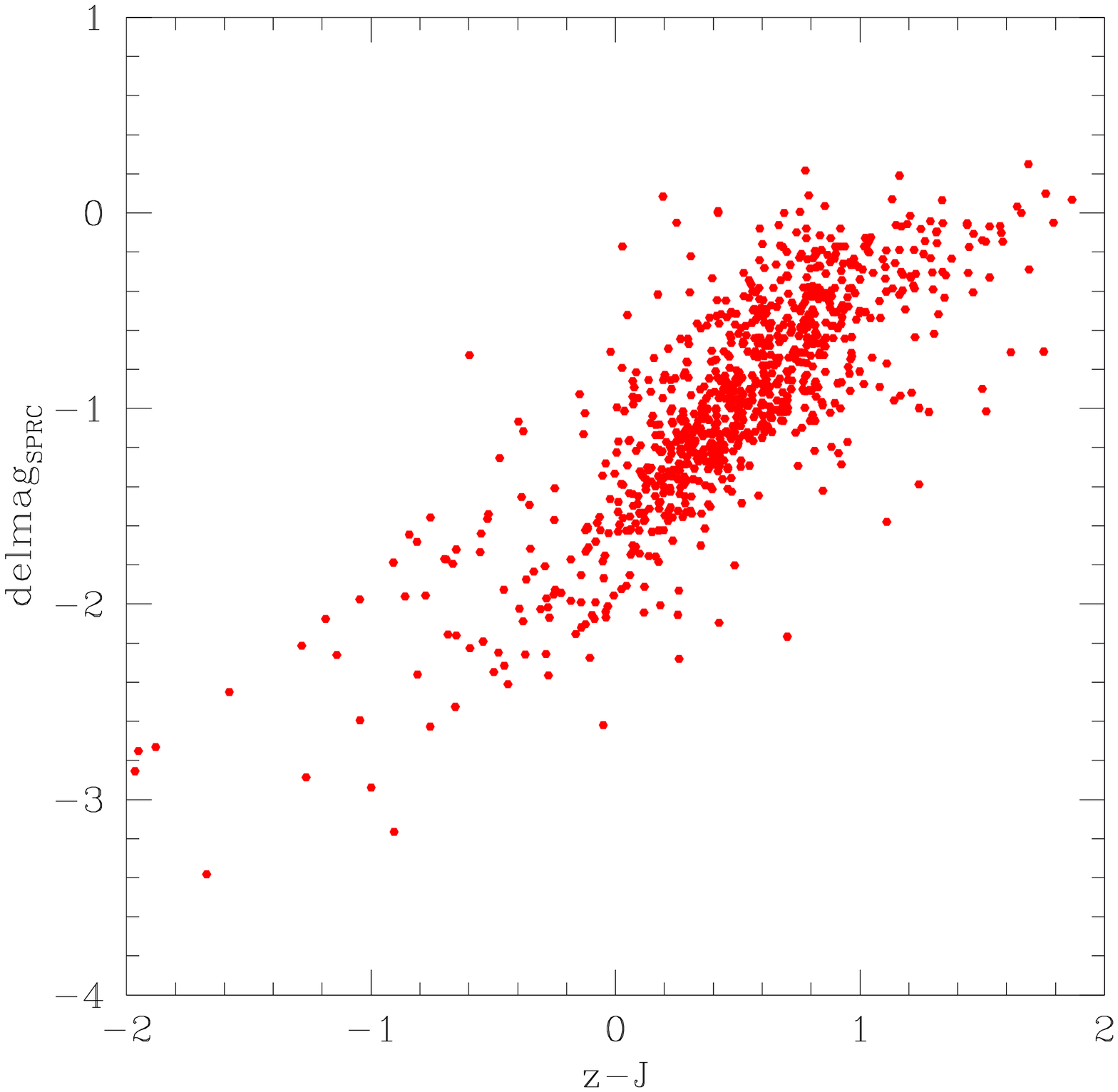,angle=0,width=7cm}
\caption{SPRC aperture correction versus z-J, with no aperture correction applied to J.
}
\end{figure}

\begin{figure*}
\epsfig{file=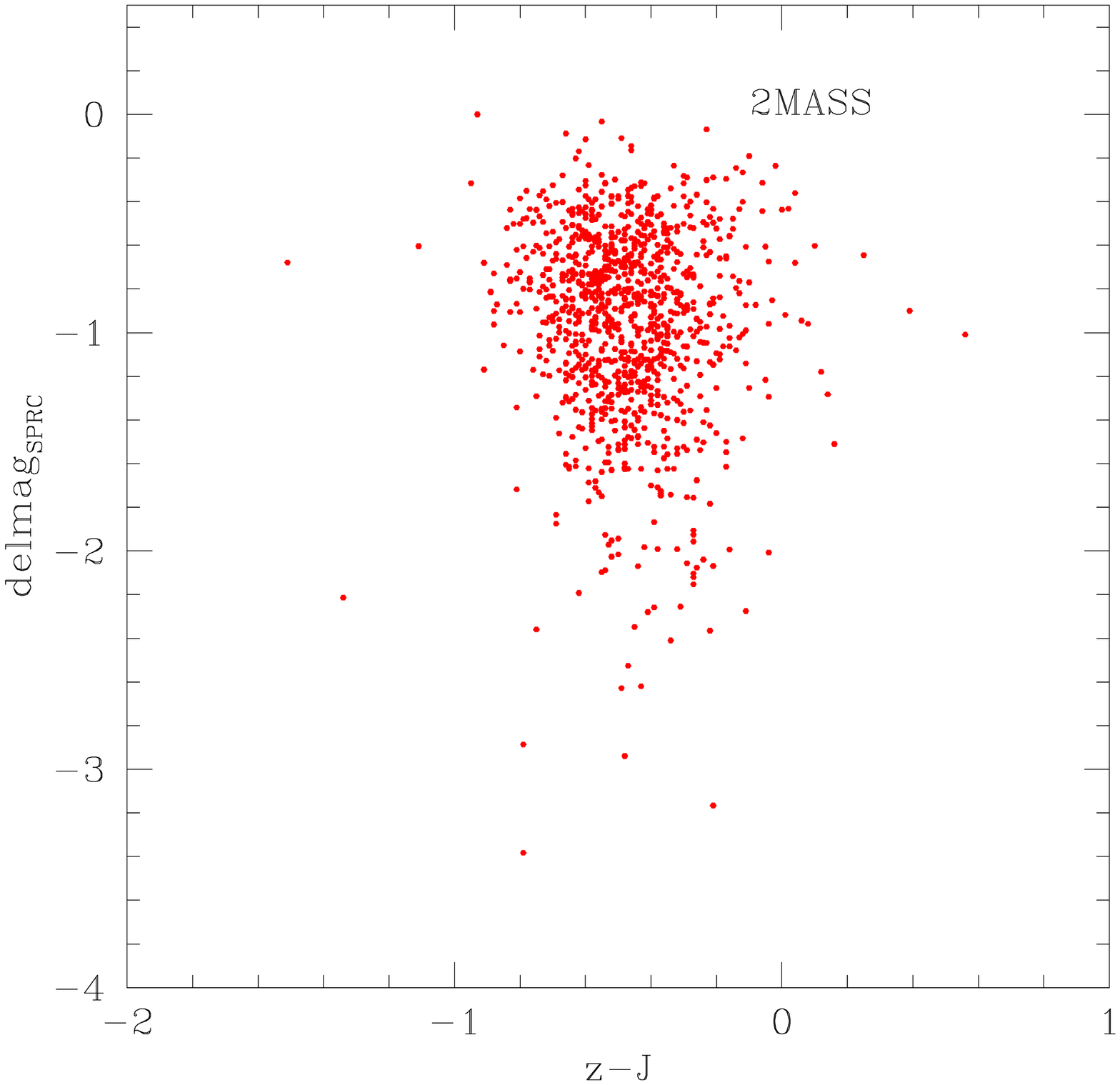,angle=0,width=7cm}
\epsfig{file=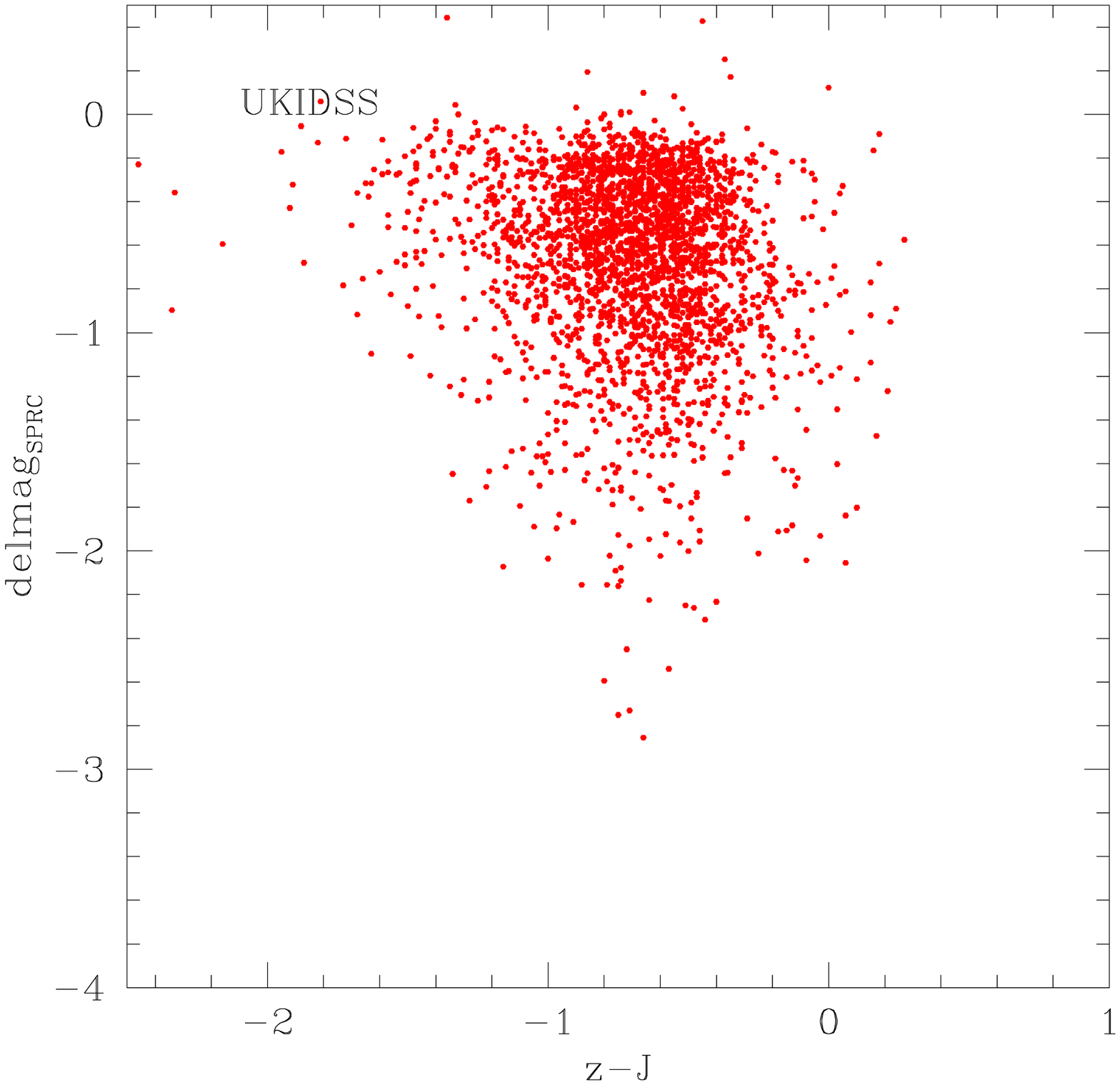,angle=0,width=7cm}
\caption{LH: SPRC aperture correction versus corrected (k=0.8) z-J for 2MASS.
RH: SPRC aperture correction versus corrected (k=1.1) z-J for UKIDSS.
Only sources with spectroscopic redshift $<$0.3 have been included to minimize the possibility of evolution of colour
with angular size.
}
\end{figure*}

\subsection{IRAC data}
In SPRC we used Kron fluxes if the 3.6 $\mu$m size was greater than a specified threshold
(area $>$ 200 pixels) , 'IRAC-aper2'
fluxes (measured in a 3.8" aperture) otherwise.  For sources smaller than this threshold, the Kron and
aper2 magnitudes agreed well.  This is also the approach adopted here.  Fig 5 shows delmag$_{SPRC}$ versus
(K-am(3.6$\mu$m)) for 2MASS (L) and UKIDSS (R) data, where am(3.6$\mu$m) is the AB magnitude derived from the Spitzer
3.6$\mu$m flux.  There are some residual issues for
the UKIDSS-IRAC comparison.   This does harm the photometric redshift estimates for some objects, resulting in a
larger reduced $\chi^2$ for the fit.  SED fits for selected galaxies with $\chi^2 > 5$ with known spectroscopic redshift shows that
the IRAC fluxes gives good consistency with the aperture corrected optical magnitudes.  The problem lies with overestimation
of the J,H, and especially K brightness for extended galaxies.  Changing the IRAC aperture correction to k*delmag$_{SPRC}$ improved the
appearance of Fig 6R enormously, but led to worse phot-z results.  The main problem was an 
offset of the $log_{10}(1+z_{ph})$ points relative to $log_{10}(1+z_{sp})$ by $\sim$0.01.  This was not 
fixed by the process of in-band correction factors (Ilbert et al 2006), and it would probably be necessary
to revise the templates in the near infrared to achieve convergence.

\begin{figure*}
\epsfig{file=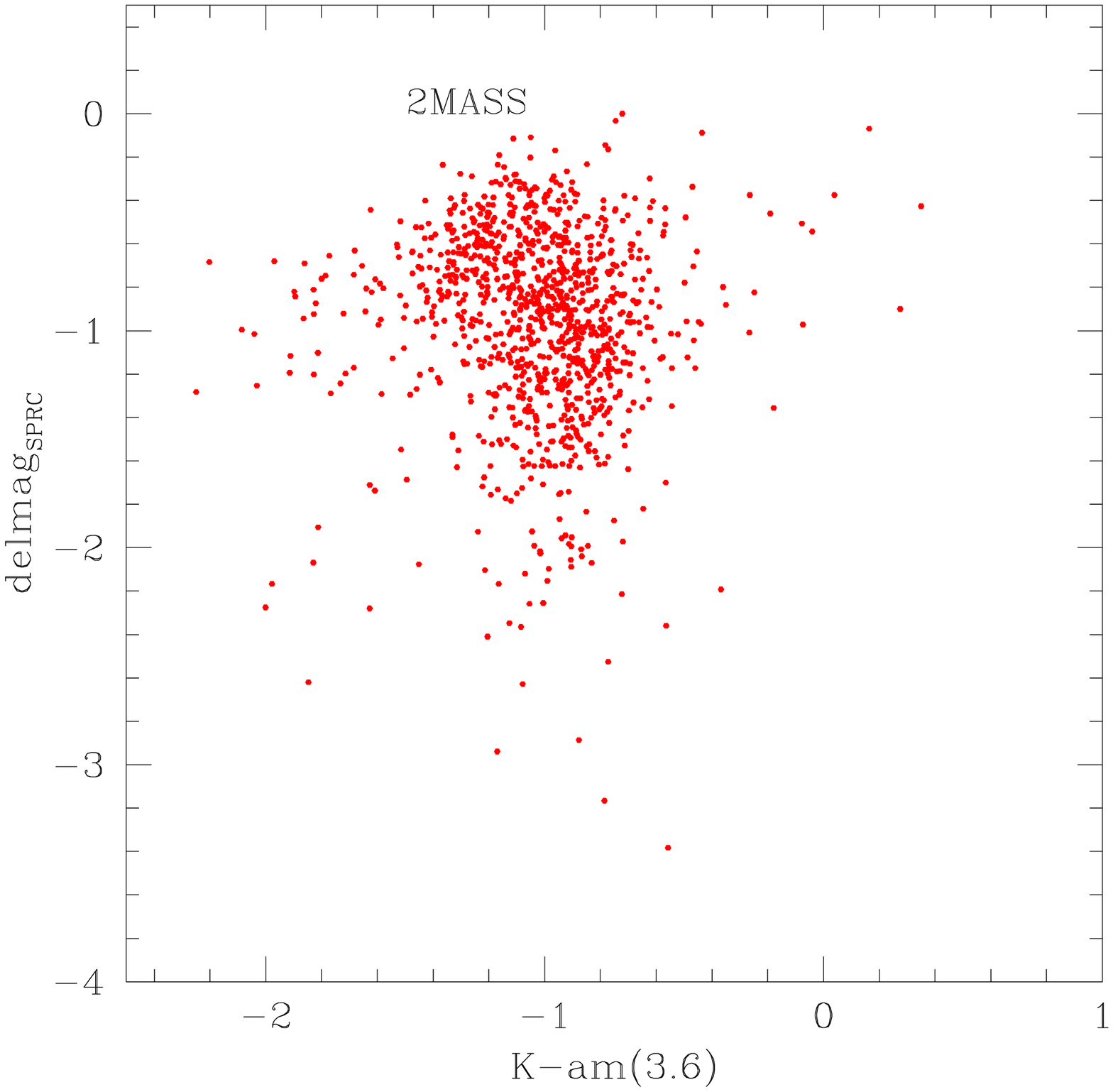,angle=0,width=7cm}
\epsfig{file=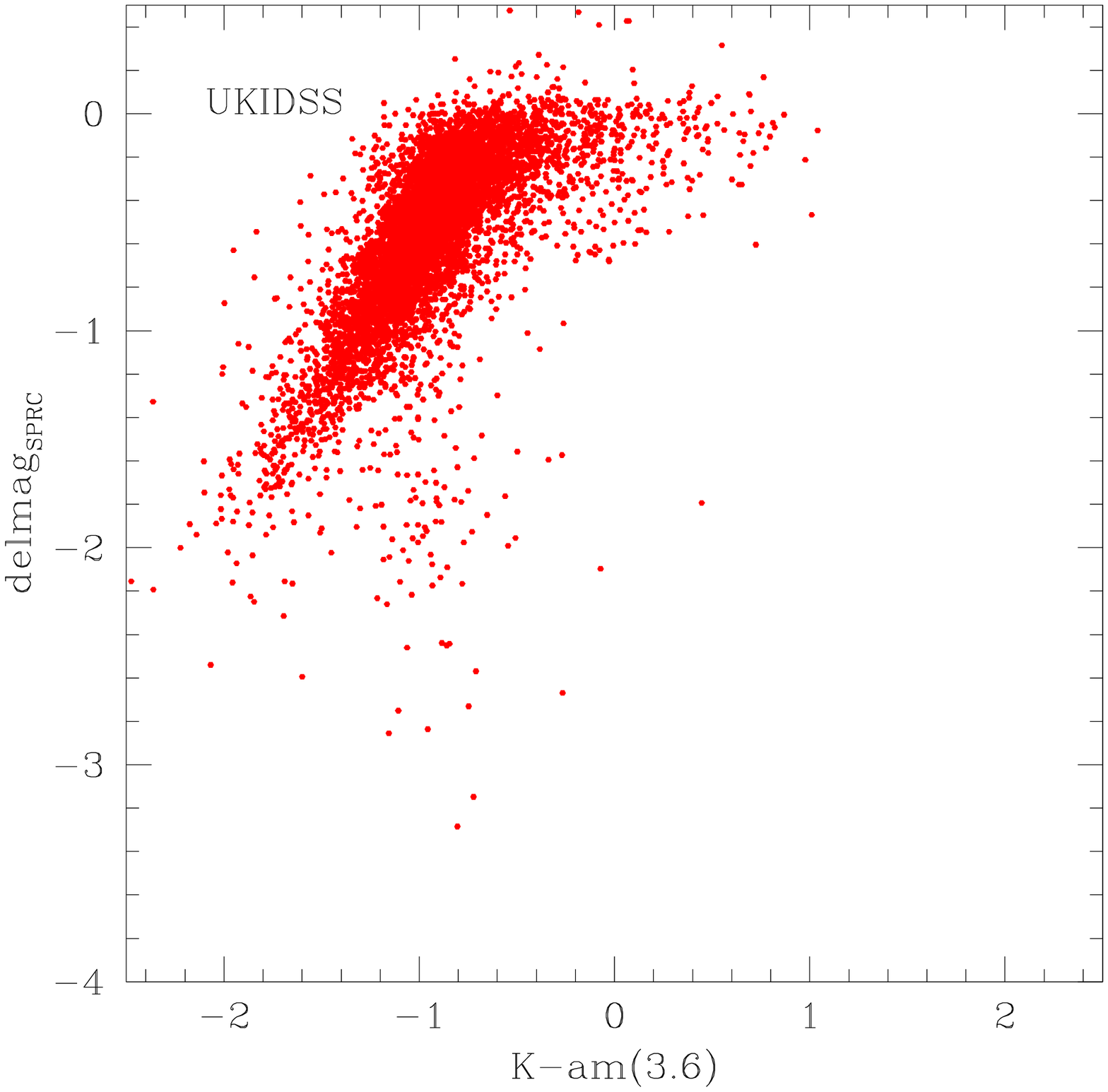,angle=0,width=7cm}
\caption{LH: SPRC aperture correction versus corrected (k=0.7) K-am(3.6) for 2MASS.
RH: SPRC aperture correction versus corrected (k=1.1) K-am(3.6) for UKIDSS.
Only sources with spectroscopic redshift $<$0.3 have been included to minimize the possibility of evolution of colour
with angular size.
}
\end{figure*}

\section{Photometric redshifts}
The photometric redshift method is as described in SPRC.  It is a two-pass template method based on 6 galaxy (
11 in the second pass) and 3 AGN templates.  The galaxy templates were first generated empirically, then modelled
with a full star-formation history so that star-formation rates and stellar masses can be derived.  After the first pass
the infrared excess relative to the starlight fit is modelled with four infrared SED types (cirrus, M82 starburst, Arp220
starburst and AGN dust torus).  Dust extinction with $A_V$ ranging up to 1.0 is permitted.  For QSOs SMC dust properties
are assumed.

For galaxies and QSOs, JHK magnitudes, and 3.6 and 4.5 $\mu$m fluxes were used in the first pass through the data provided 
there was no evidence of a strong AGN dust torus component, determined by condition  S(5.8) $>$ 1.2 S(3.6).
For galaxies these near infrared data were not used in the second pass fit if the
dust torus component  was dominant at 8 $\mu$m from the ir template fitting.  For the second pass for QSOs, a new
approach was developed, described in section 4 below, to allow use of the near infrared data even though a dust torus 
component is present.

In the photometric redshift solution
we used the new SDSS 'model' magnitudes, and the revised WFC magnitudes, but we retained the WFC star/galaxy 
classification in each band (because this information is available for a much higher proportion of sources) and the 
optical aperture correction used in SPRC.  We used 2MASS JHK (PSC), where
available, and UKIDSS JK (aper3) if not.  

We used galaxies with known spectroscopic redshifts to determine in-band correction factors, following
Ilbert et al (2006).   These are shown in Table 2.  Spectroscopic redshifts came from a wide range of references
(details are given in the on-line catalogue) and are mostly in the range 0-1.2 for galaxies.

\begin{table*}
\caption{Correction factors for fluxes in each band, by area}
\begin{tabular}{lllllllllll}
area & U' & g' & r' & i' & Z' &  J & H & K & 3.6 $\mu$m & 4.5 $\mu$m\\
&&&&&&&&&\\
EN1,EN2 & 1.354 & 1.135 & 1.104 & 1.059 & 0.650 &  0.976 & 0.900 & 0.888 & 1.071 & 1.104\\
Lockman & 1.360 & 1.062 & 0.982 & 0.922 & & 0.923 & 0.904 & 0.899 & 1.036 & 1.066\\
VVDS & 1.046 & 0.969 & 0.928 & 0.960 & 0.972  & 1.173 & & 1.029 & 1.085 & 1.015\\
XMM-LSS & 1.308 & 1.206 & 1.056 & 0.925 & 1.011 & &&& 1.085 & 1.086\\
CDFS & 1.046 & 0.969 & 0.916 & 0.991 & 0.972 & &&& 1.238 & 1.037\\
\end{tabular}
\end{table*}

\begin{figure*}
\epsfig{file=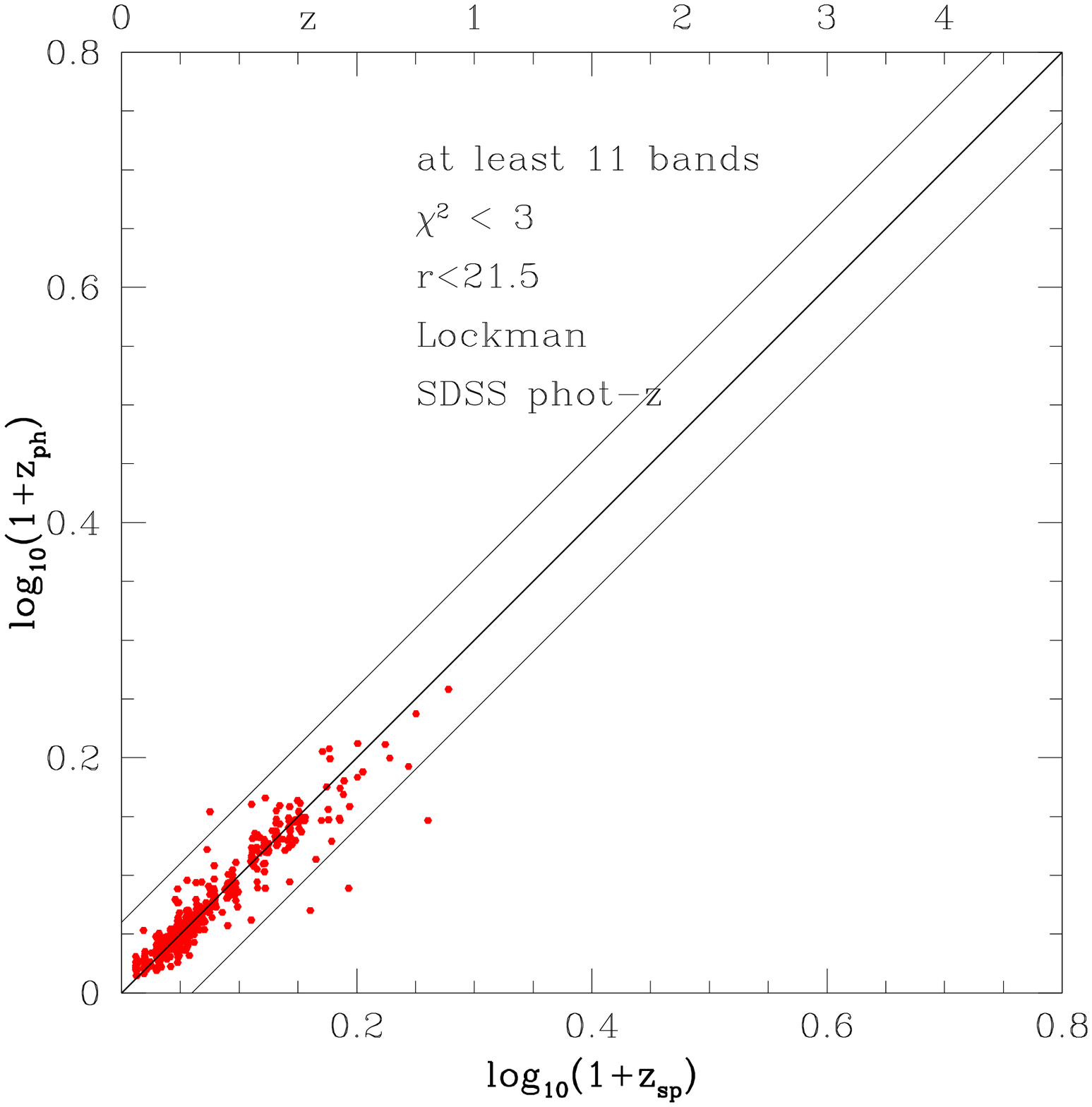,angle=0,width=7cm}
\epsfig{file=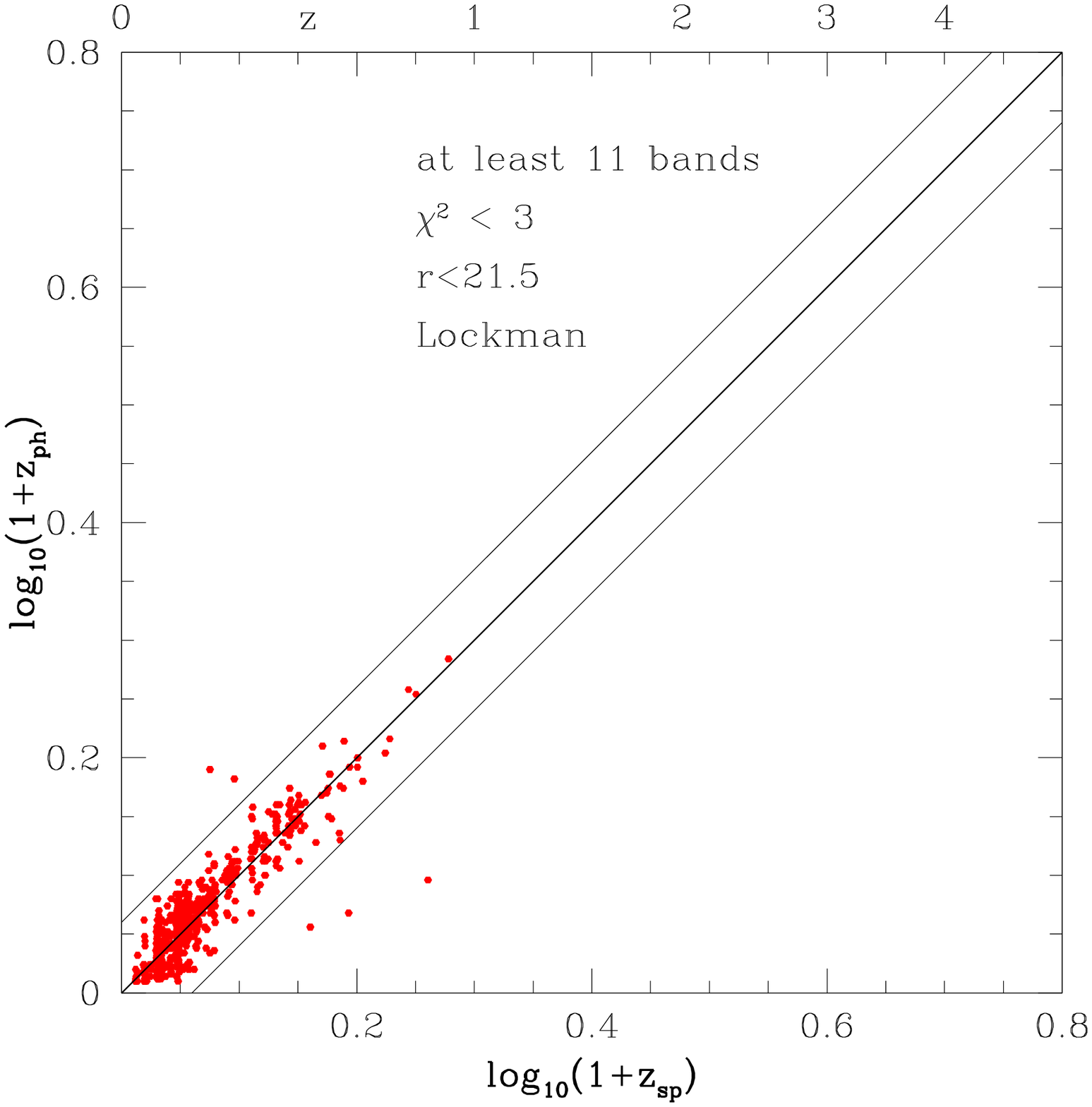,angle=0,width=7cm}
\caption{LH: SDSS photometric redshifts versus spectroscopic redshifts for Lockman SWIRE sample.
RH: photometric redshifts from present work versus spectroscopic redshifts for Lockman SWIRE sample.
}
\end{figure*} 

Fig 6 shows a comparison of the SDSS $log_{10} (1+z_{phot})$ with  $log_{10} (1+z_{spect})$ for the SWIRE
Lockman sample (LH). and the same plot for the present sample (RH).  The SDSS performance is better
at z $<$ 0.3, but our approach works better at z $>$ 0.5.  To improve our photometric redshifts at z $<$ 0.3
it would be necessary to refine our optical templates using the new photometric data, and to increase the 
number of templates.  The latter would be likely to worsen the performance at higher redshift through 
increased aliasing.

Fig 7L shows the same comparison for the SWIRE Photometric Redshift Catalogue, restricted to r $<$ 23.5, 
reduced $\chi^2 <3$ and at least 5 photometric bands in the solution.   Fig 7R shows a similar plot for
the revised SPRC, with the requirement that K be selected, and for a minimum of 7 photometric bands.
We can see that inclusion of near infrared (JHK) data in the solution has reduced the number of
catastrophic outliers.

Fig 8L shows the same comparison for the revised catalogue in XMM, with at least 9 bands in the solution
in the solution.  Fig 8R shows the same comparison for photometric redshifts from the LePhare
method (Ilbert et al 2006).  The latter results are better for z $<$ 0.5, but worse for z $>$ 0.5.  Again this
is probably the consequence of the larger number of templates used by Ilbert et al.

\begin{figure*}
\epsfig{file=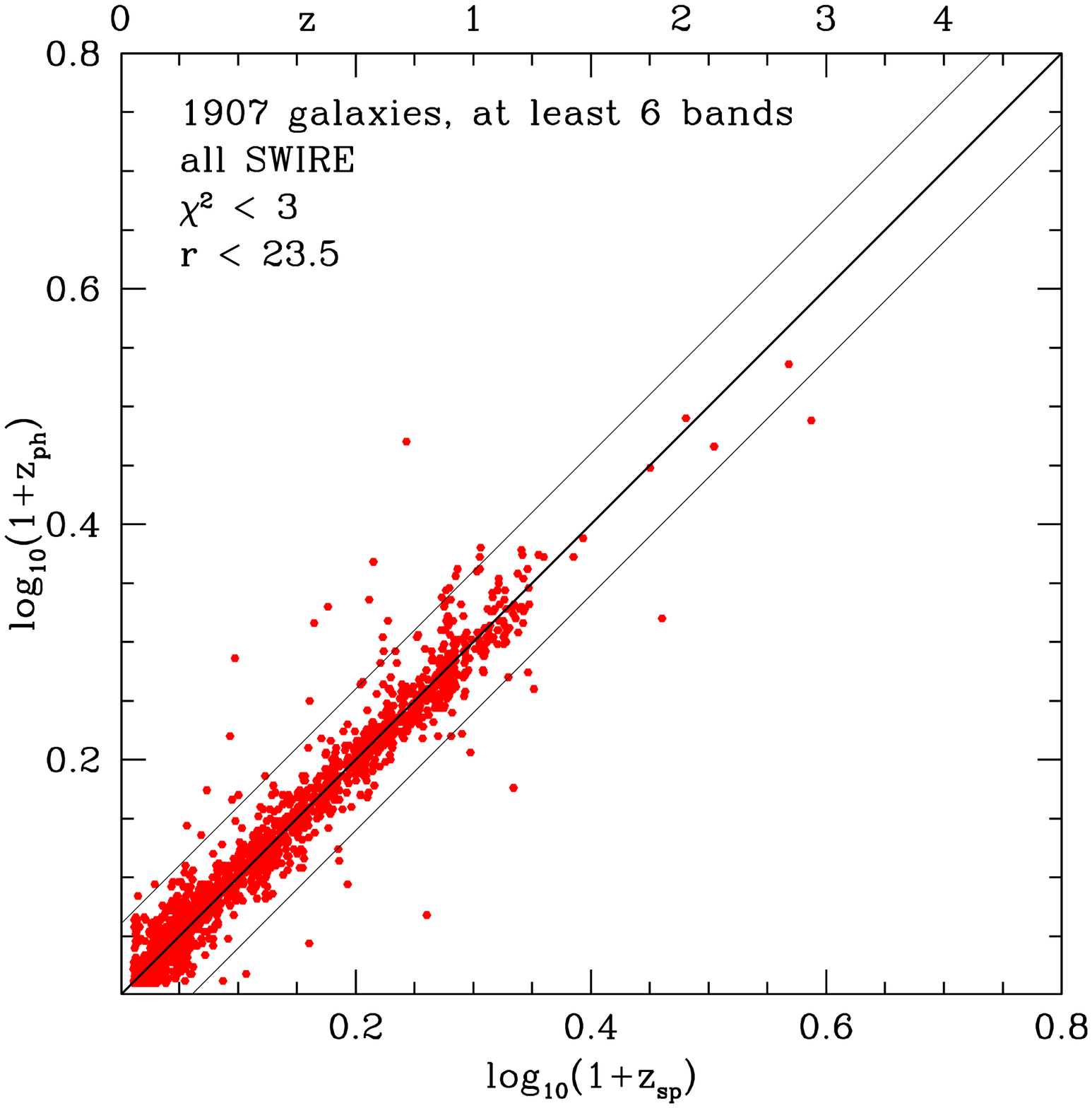,angle=0,width=7cm}
\epsfig{file=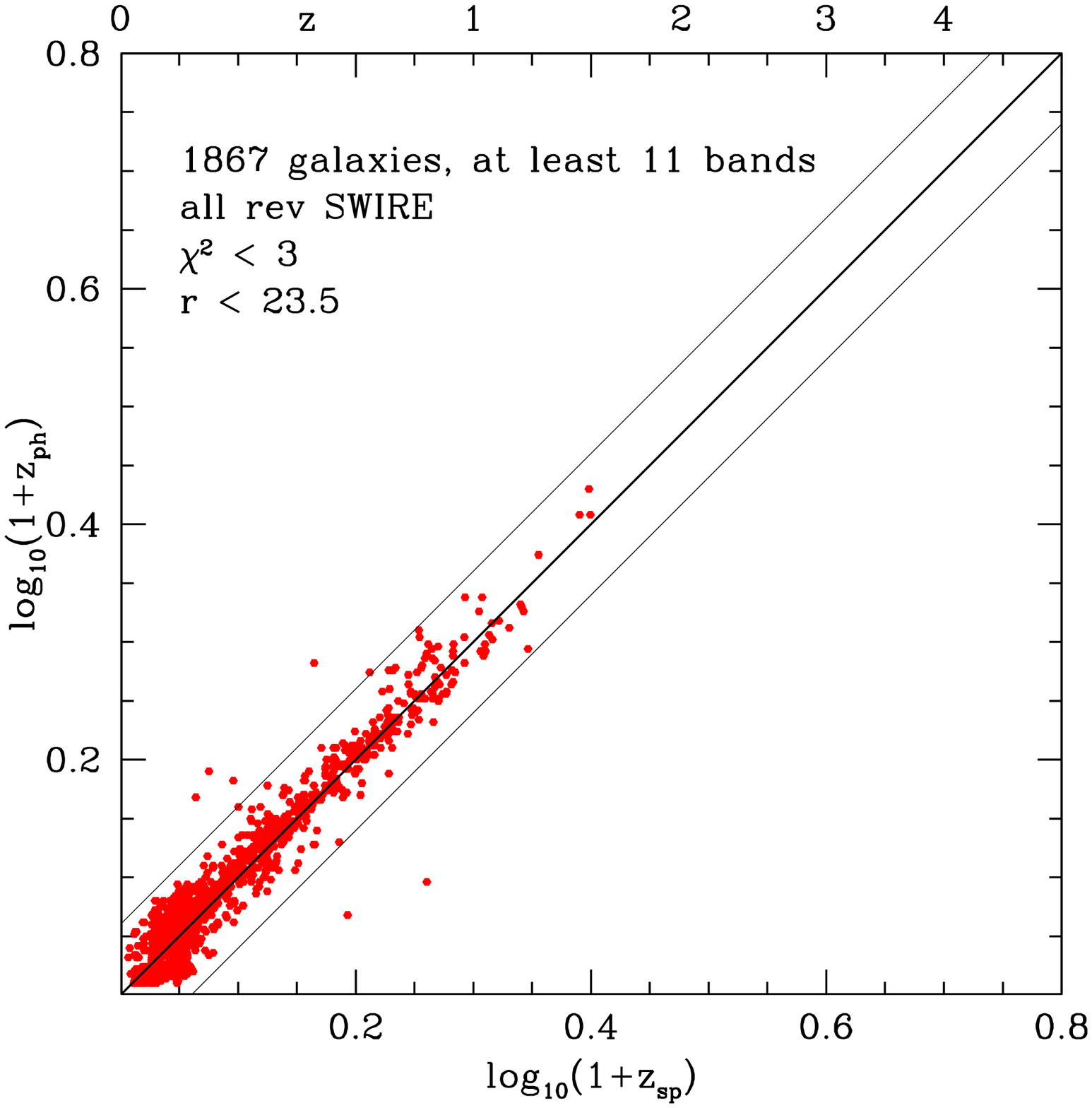,angle=0,width=7cm}
\caption{LH: Photometric redshifts  from SWIRE Photometric Redshift Catalogue (Rowan-Robinson et al 2008).
for galaxies with r $<$ 23.5, at least 6 photometric bands, reduced $\chi^2 < 3$.  Red symbols: galaxies.
RH: Same for revised SPRC, with at least 11 photometric bands.
}
\end{figure*}

\begin{figure*}
\epsfig{file=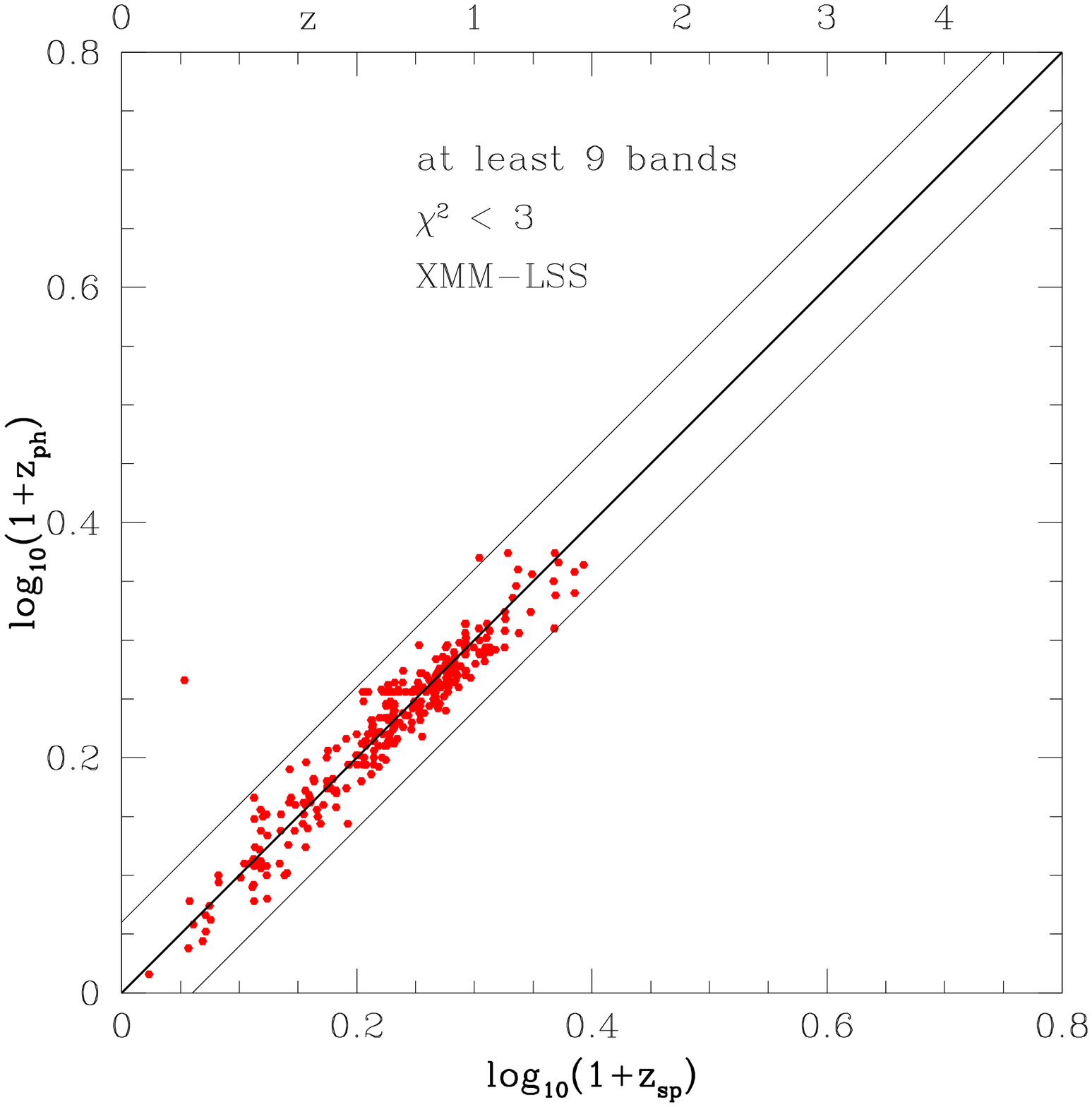,angle=0,width=7cm}
\epsfig{file=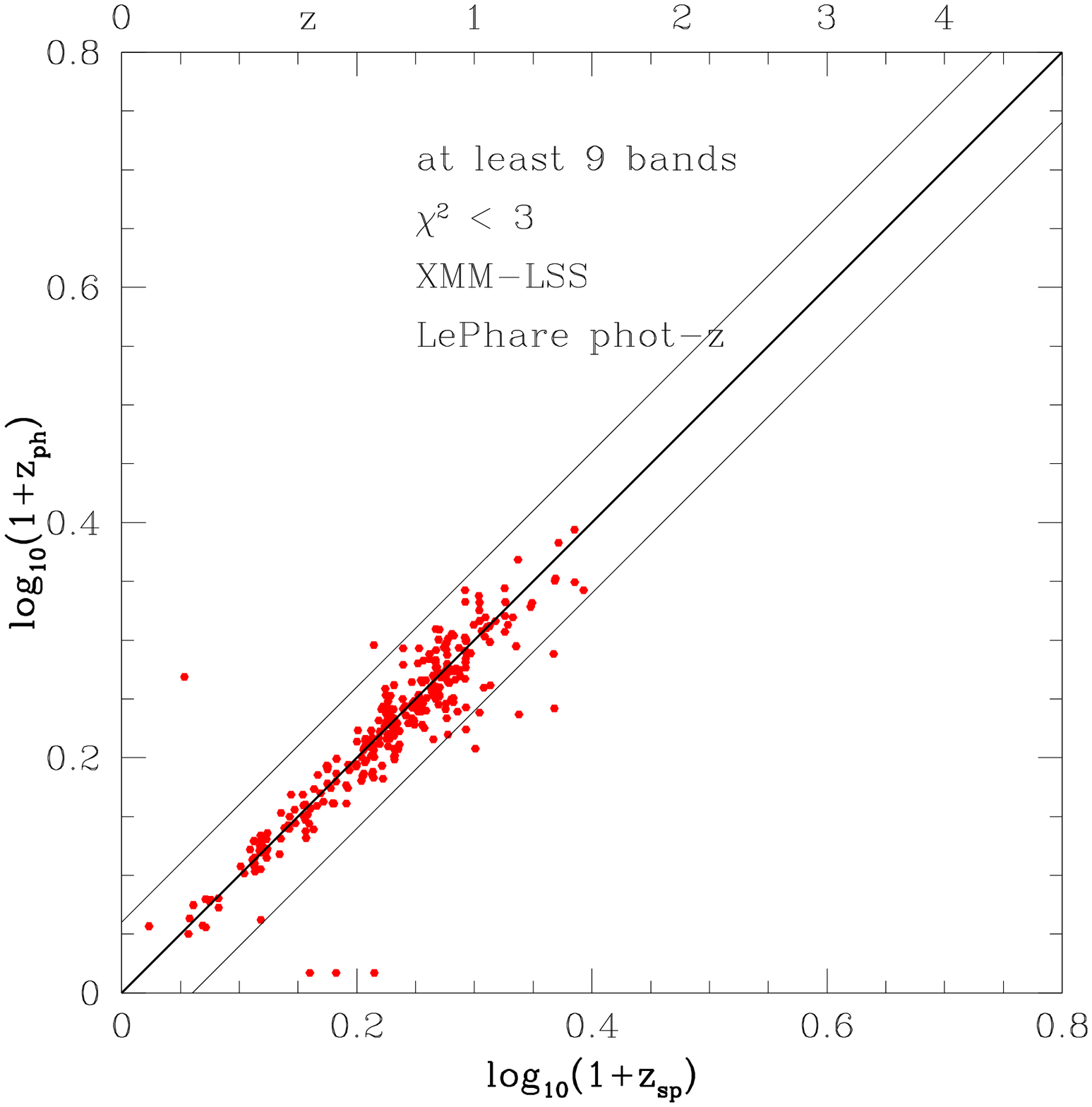,angle=0,width=7cm}
\caption{L: Photometric redshifts in XMM-LSS using new fusion catalogue: ugriz from SDSS, revised ugri from WFC,  JHK from 2MASS, JK from UKIDSS. R: Photometric redshifts from LePhare method (Ilbert et al 2009).
}
\end{figure*}


\section{QSOs}
The SPRC approach required that an object be flagged as stellar to consider a QSO optical template.  With the SPRC stellar flag
some QSOs get missed (and end up with the wrong redshift) as a result of this condition.  As discussed in SPRC
it is not possible to allow a QSO template option for all galaxies, since far too many galaxies end up with
mistakenly high redshifts.  However, while still requiring a stellar flag for an object to select a QSO
 template, we have allowed the SDSS stellar flag to override the WFC flag where they 
disagree and this allows a few more quasars through. 

Salvato et al (2009) have demonstrated excellent performance for 1032 QSOs and AGN in the COSMOS field, using 30 photometric bands, 
including 12 narrow-band filters.  They introduce two innovations: firstly they track the variabilty of quasars and apply
an appropriate correction to the photometry.  Secondly they use templates that include a range of contributions from AGN dust
tori.

\begin{figure*}
\epsfig{file=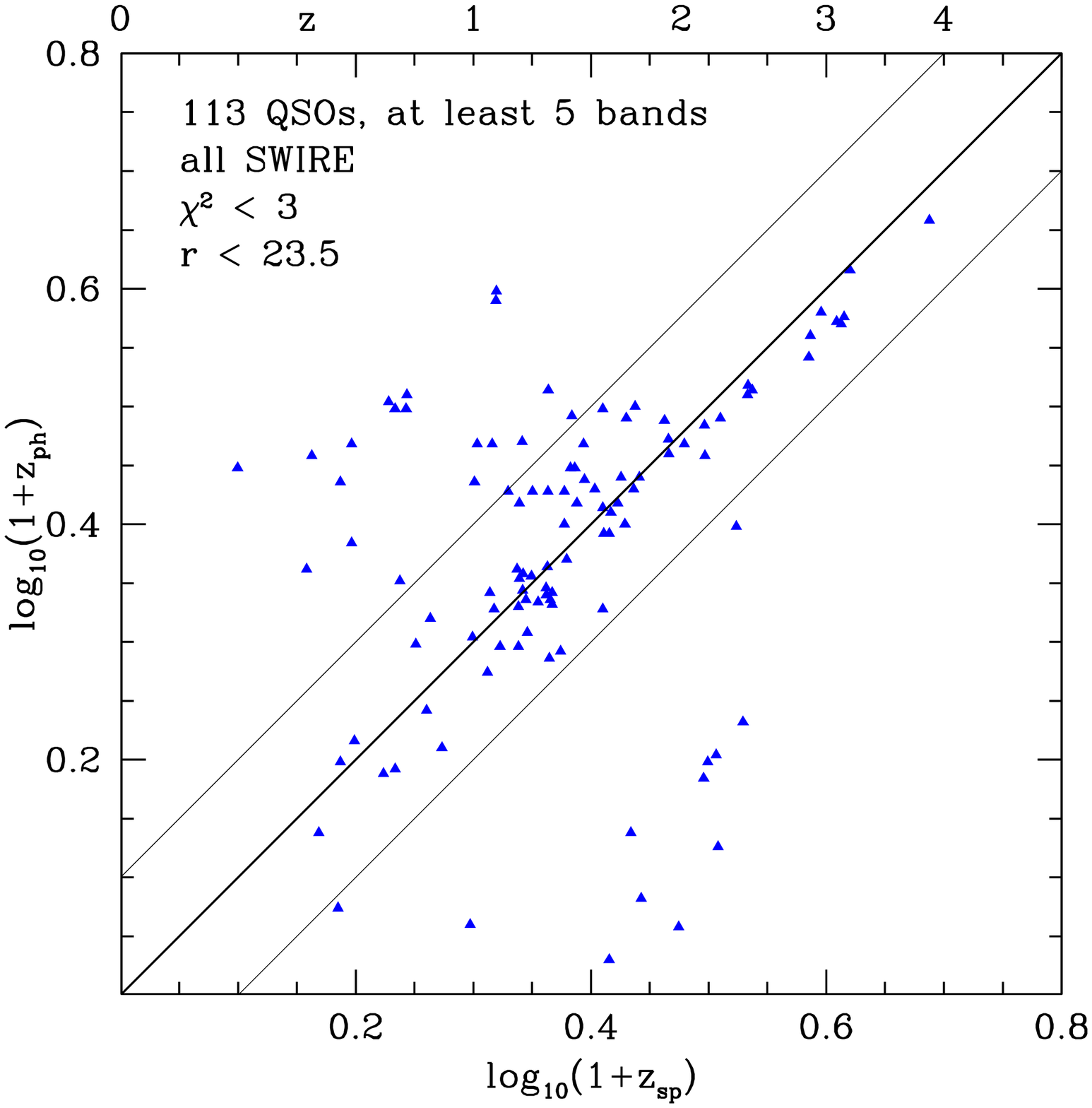,angle=0,width=7cm}
\epsfig{file=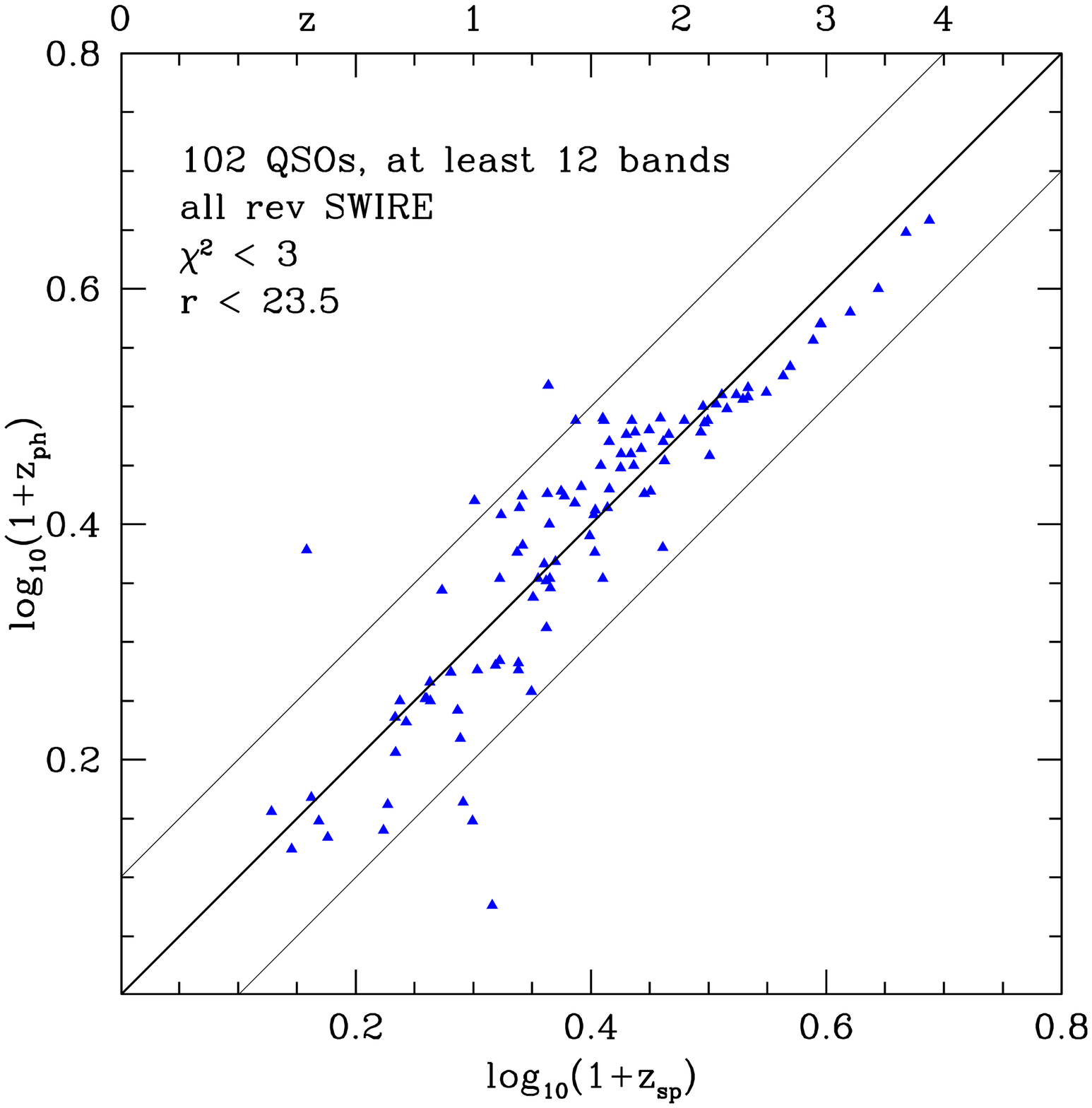,angle=0,width=7cm}
\caption{LH: Photometric redshifts  from SWIRE Photometric Redshift Catalogue (Rowan-Robinson et al 2008).
for quasars with r $<$ 23.5, at least 7 photometric bands, reduced $\chi^2 < 3$.  Blue symbols: QSOs.
RH: Same for revised SPRC, with at least 12 photometric bands.
}
\end{figure*}

While we do not have the information to track QSO variability in our photometry, we have explored the idea of adding
a range of AGN dust tori strengths to our QSO templates during the second pass, and then using the 1.25-8 $\mu$ m data in the redshift solution.
The amplitude of dust tori added corresponded to $L_{tor}/L_{opt}$ = 0, 0.2, 0.4, 0.6, 0.8, 1.0.  
The performance for QSOs is much improved (Fig 9), especially in the reduction of catastrophic outliers.  For at least 11 photometric bands,
reduced $\chi^2 < 3, r < 21.5$, we find an rms of 9.3$\%$ and an outlier rate 9.3$\%$.  For comparison, Salvato et al 
(2009), with 30 bands, achieved an rms of 1.2$\%$ for QSOs with $I<22.5$, and an outlier rate of 6.3$\%$.  Their greatly
improved rms can be attributed to the correction for variability and to the use of 30 photometric bands, including 12 narrow 
band filters.  But our outlier performance is almost as good, despite less than half the number of photometric
bands.  

\section{Reasons for poor $\chi^2$}

Although the photometric redshift estimates are good for sources where the reduced $\chi^2 < 3$, the estimates get
slightly worse as $\chi^2$ increases and some sources have very large $\chi^2$.  We have investigated the
reasons for poor $\chi^2$ through colour-colour diagrams and SED plots.  The main reasons found for poor
$\chi^2$ are found to be (i) contamination by stars, (ii) poor photometry, (iii) QSOs misclassified as galaxies, (iv) QSOs for 
which the extinction is higher than the maximum assumed $A_V$ = 1.

\begin{figure*}
\epsfig{file=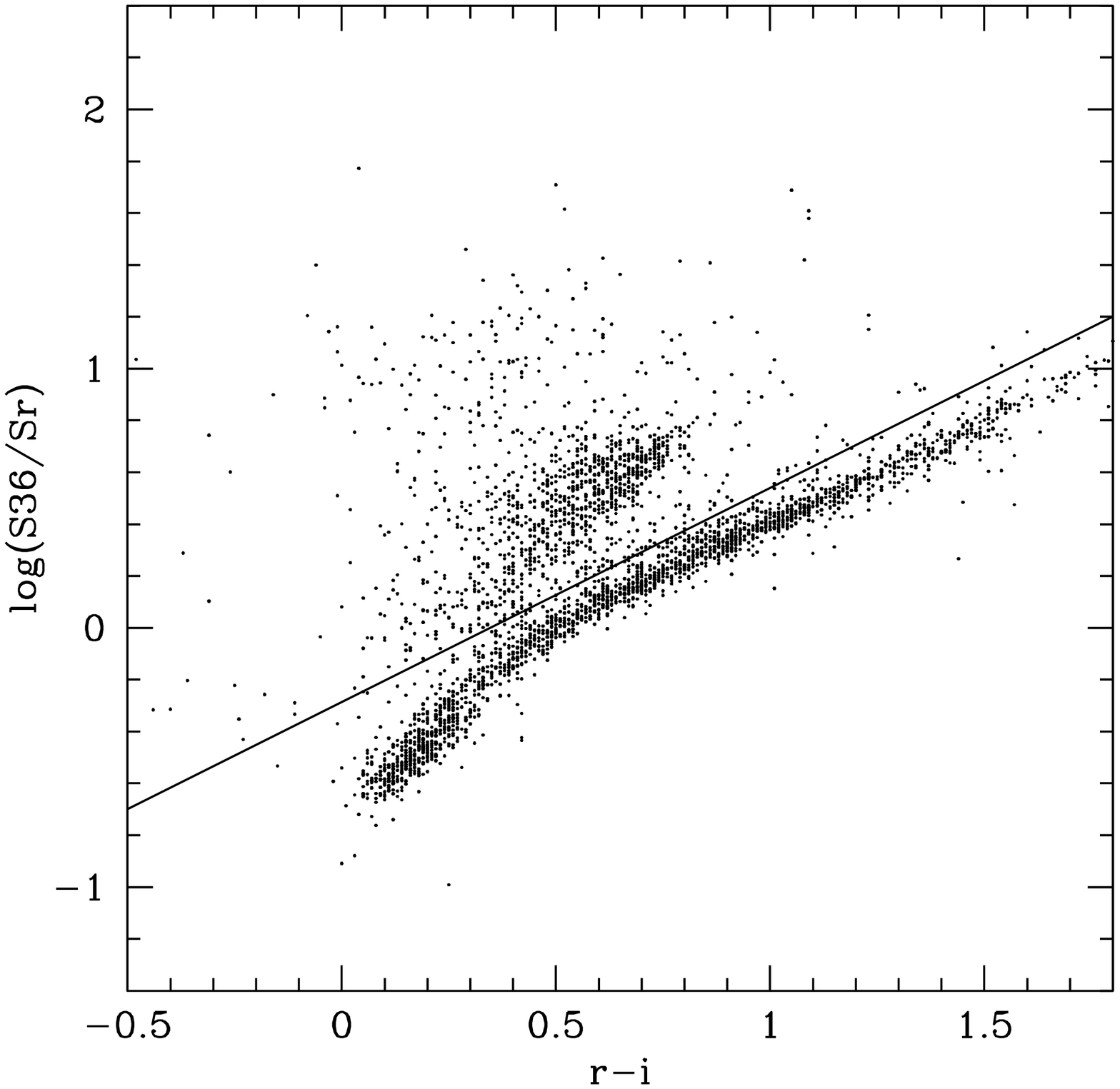,angle=0,width=7cm}
\epsfig{file=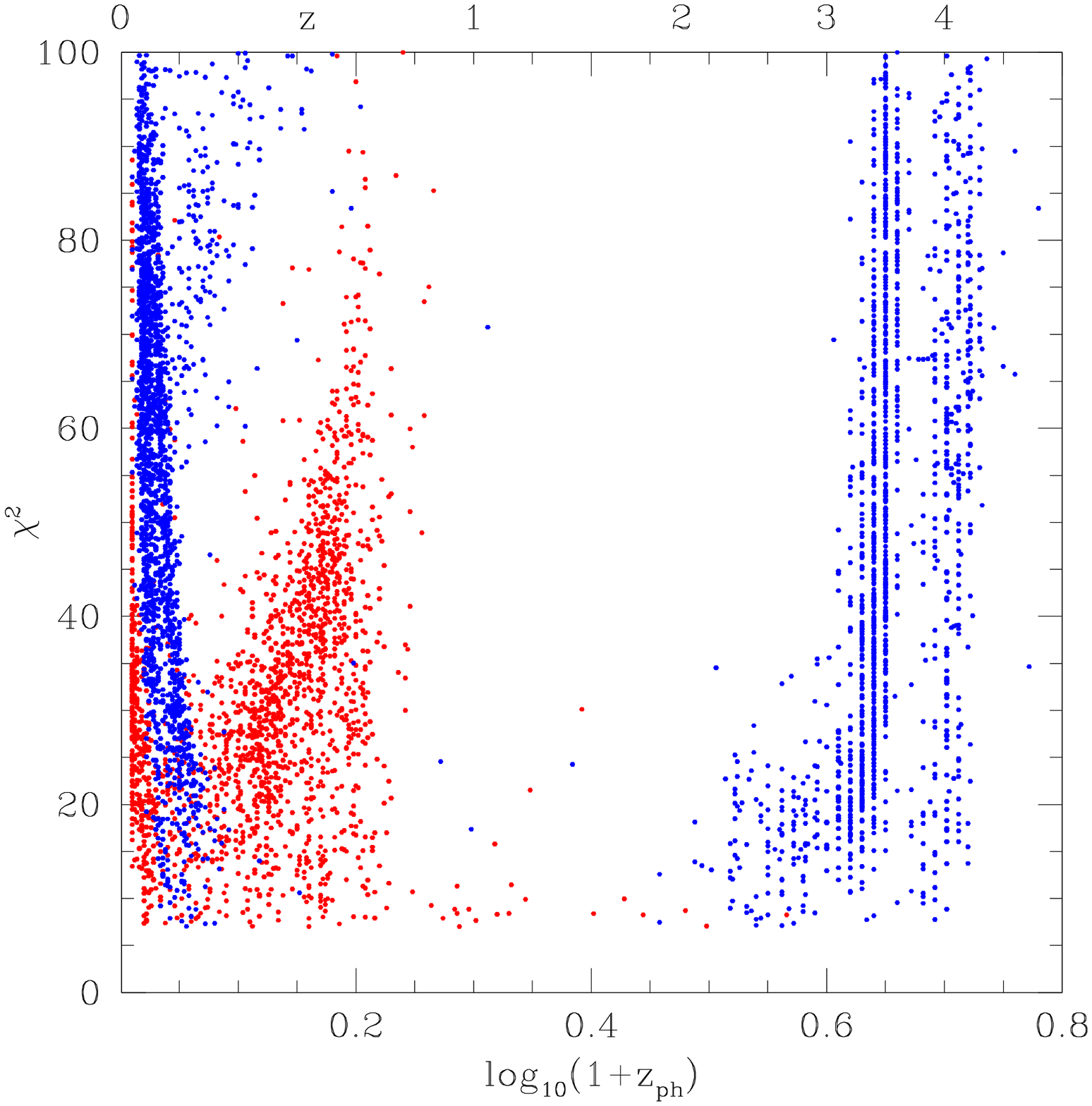,angle=0,width=7cm}
\caption{LH: Ratio of 3.6 $\mu$m to r-band flux versus r-i for whole catalogue, for sources with $\chi^2 > 20$ and at least 12 photometric
bands.  A clear stellar sequence is seen.  Stars have been excluded from the catalogue by removing objects flagged as
point sources lying below the solid line.
RH: Reduced $\chi^2$ versus estimated photometric redshift for sources classified as stars.  Red symbols:
sources which the photometric redshift code identified as galaxies; blue symbols: QSOs.
}
\end{figure*}

Stars can be recognised in a plot of the ratio of 3.6 $\mu$m to r-band flux versus r-i ( Fig 10L, see Rowan-Robinson 
et al 2005).  A clear stellar sequence can be seen.  We have removed stars from the catalogue by excluding
17326 sources with stellar flags which lie below the solid line in Fug 10L, and which had reduced $\chi^2 > 7$.  Figure 10R shows 
the $\chi^2$ and photometric redshift generated for these candidate stars by the photometric code.  Stars
can generate spectacularly bad $\chi^2$ and tend to alias with specific galaxy and QSO redshifts.  We believe
that few of the sources with $\chi^2 < 7$ are stars.

\begin{figure*}
\epsfig{file=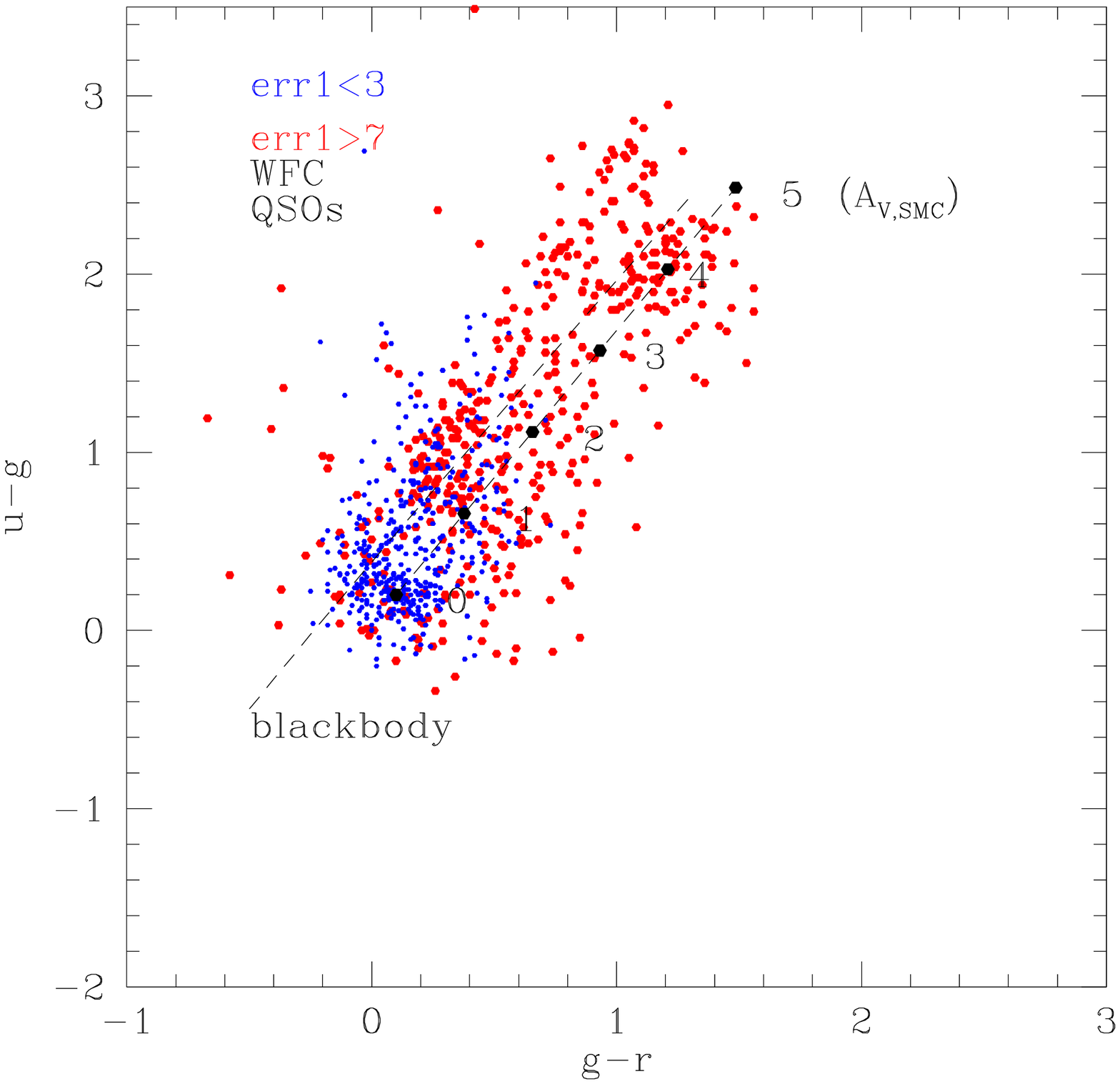,angle=0,width=7cm}
\epsfig{file=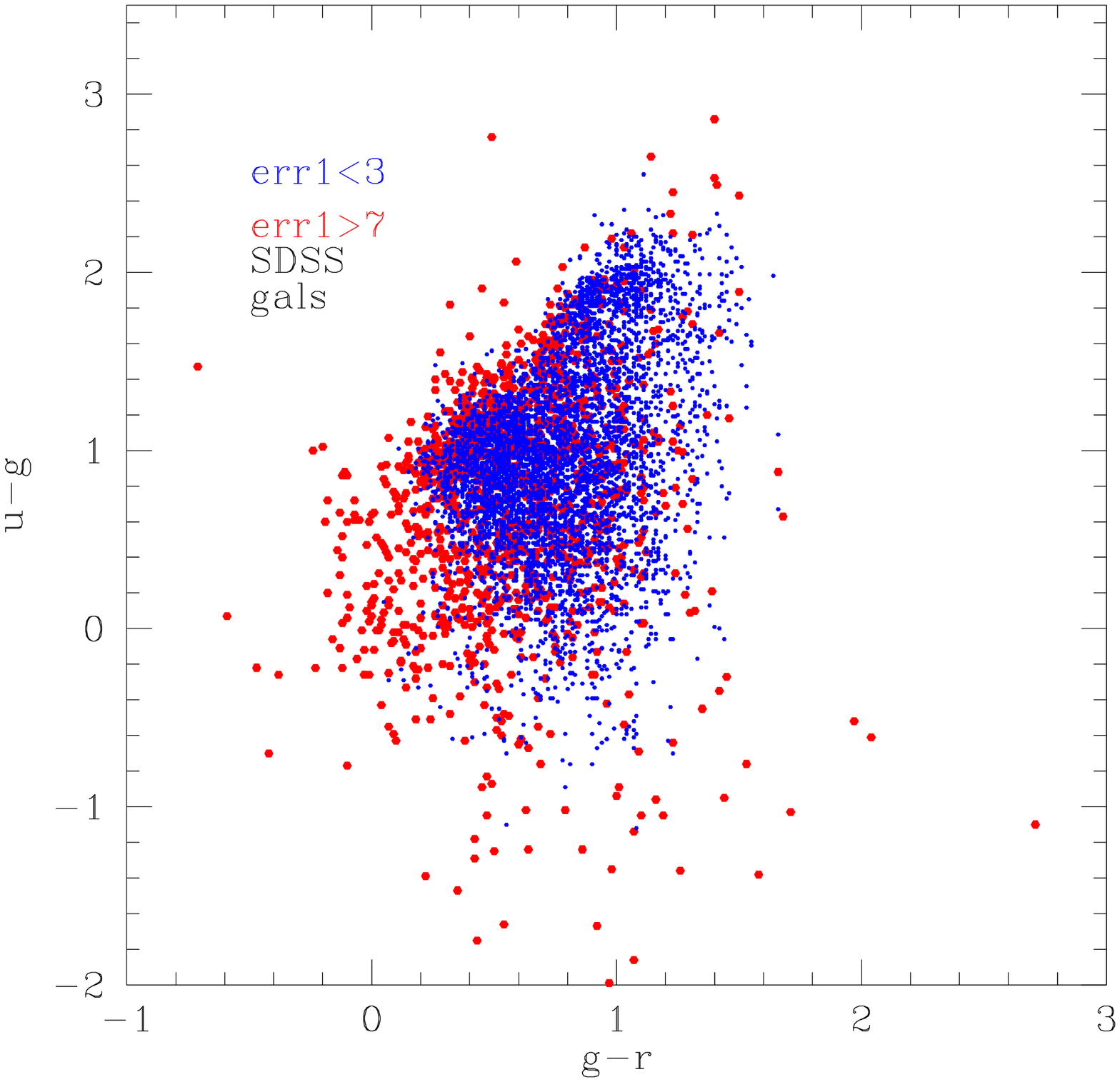,angle=0,width=7cm}
\caption{LH: (u-g) versus (g-r) (WFC photometry) for QSOs in EN1 field with at least 11 photometric bands, 
with blue symbols for sources with $\chi^2 < 3$
and red symbols for $\chi^2 > 7$, after removal of stars using the 3.6$\mu$m-(r-i) constraint.  The broken
locus is for blackbodies with temperature ranging from 3000 to 15000, the solid locus shows the effect
of extinction ($A_V$=0-5) assuming SMC dust properties.
RH: Reduced $\chi^2$ versus estimated photometric redshift for sources classified as stars.  Red symbols:
sources which the photometric redshift code identified as galaxies; blue symbols: QSOs.
}
\end{figure*}

Figure 11L shows (u-g) versus (g-r) (WFC photometry) for QSOs, with blue symbols for sources with $\chi^2 < 3$
and red symbols for $\chi^2 > 7$, after removal of stars using the 3.6$\mu$m-(r-i) constraint.  The broken
locus is for blackbodies with temperature ranging from 3000 to 15000, the solid locus shows the effect
of extinction using SMC dust properties.  The $\chi^2 > 7$ objects show a much wider range of colours than
those with $\chi^2 < 3$.  Examination of SED plots shows that most of the $\chi^2 > 7$ objects are stars
with slightly higher 3.6 $\mu$m fluxes than the stellar sequence in Fig 10L.   A few are QSOs with higher
extinction than the maximum $A_V$ = 1 we have used.  Others have high $\chi^2$ simply because of poor
photometry or SED shapes differing from our templates.

Figure 11R shows (u-g) versus (g-r), for galaxies with SDSS photometry, with different symbols for $\chi^2 < 3$ 
and $> 7$.  About 10$\%$ of 'galaxies' with $\chi^2 > 7$ achieved much improved fits with QSO templates
(some of these did not have the stellar flag that was a prerequisite for consideration as a QSO).  Some are
stars with slightly higher 3.6 $\mu$m fluxes than the stellar sequence in Fig 10L.   For the rest the problem
appears to be poor photometry (and underestimated photometric errors).

\begin{figure*}
\epsfig{file=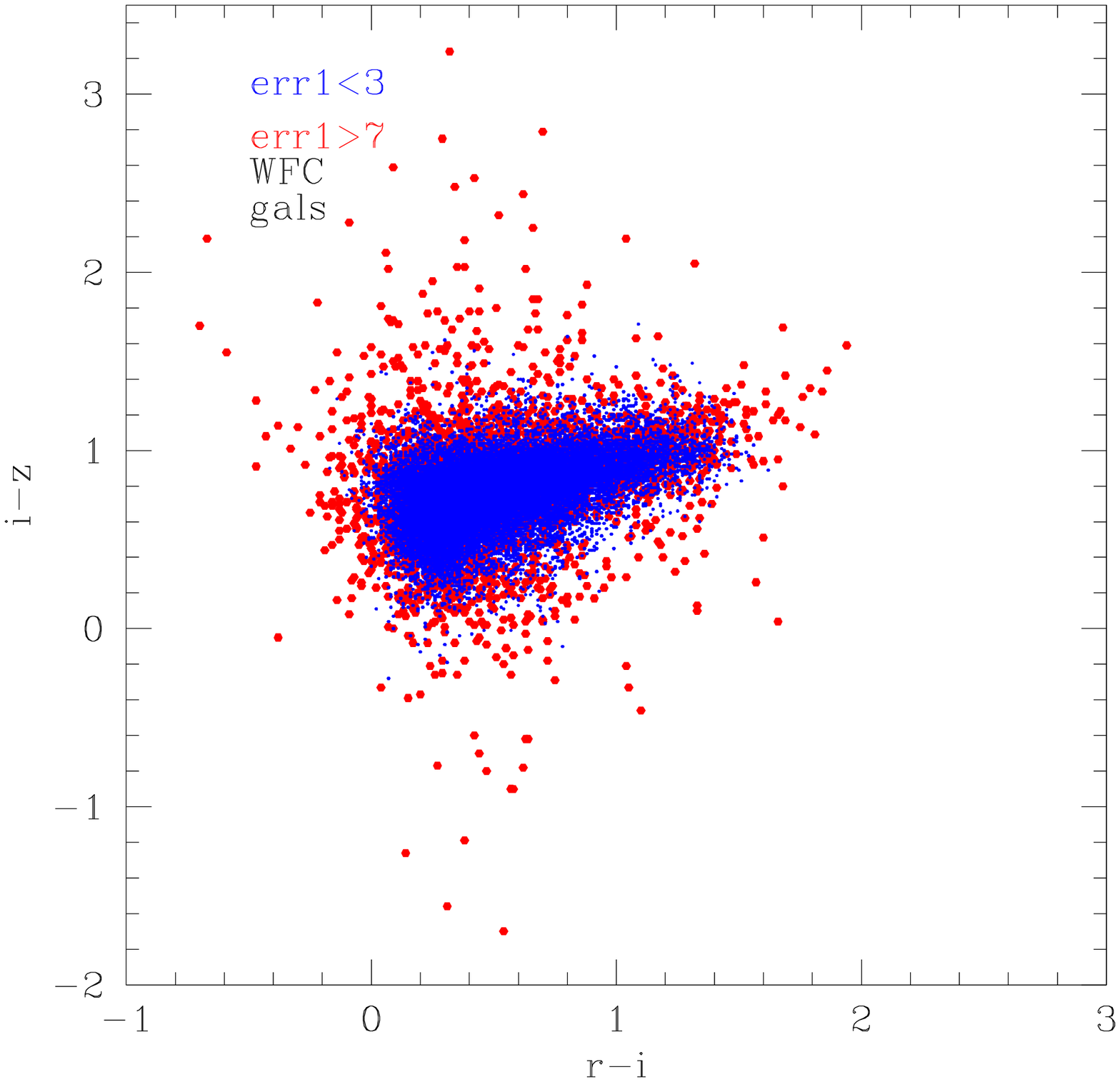,angle=0,width=7cm}
\epsfig{file=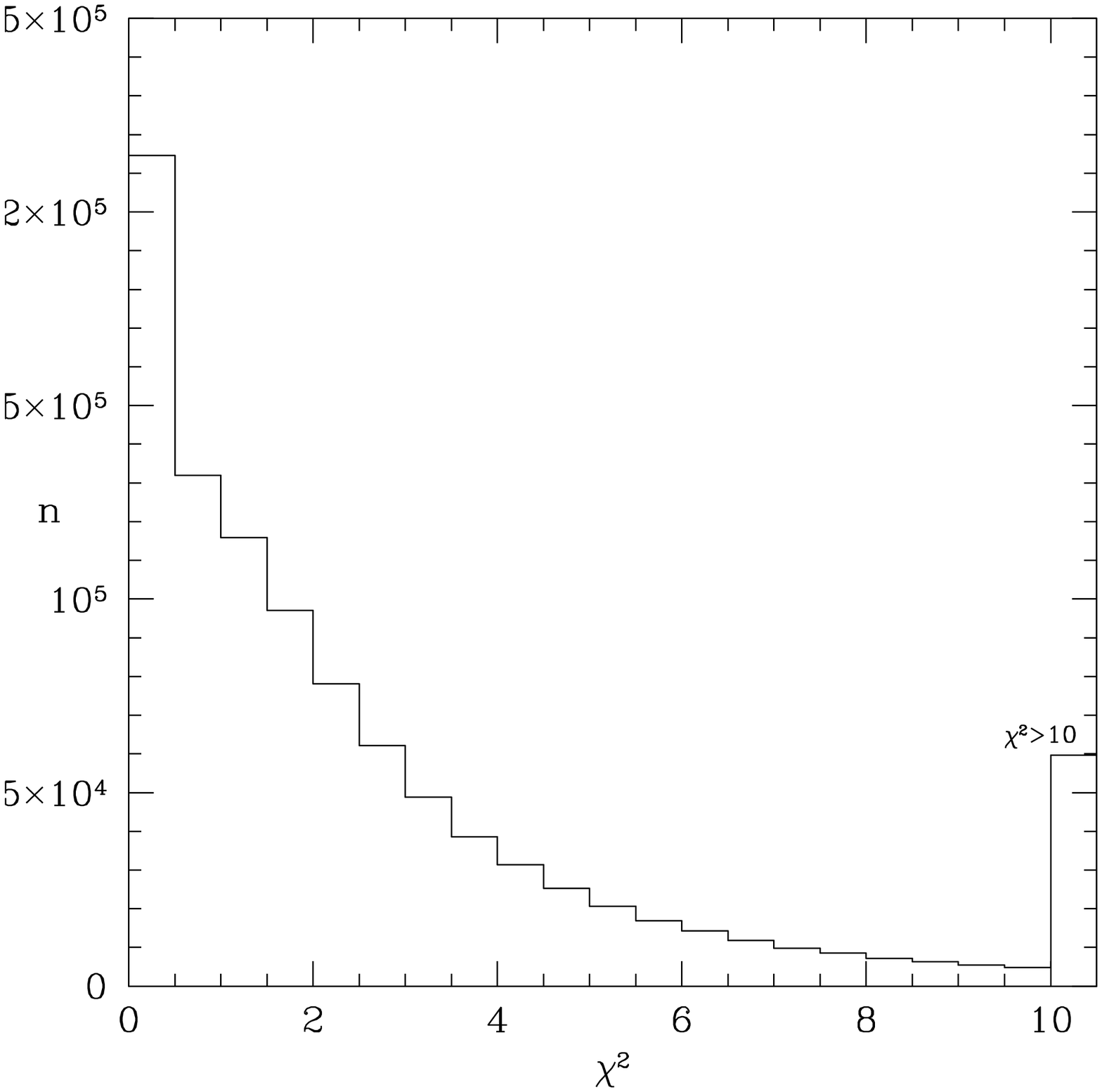,angle=0,width=7cm}
\caption{L: (i-z) versus (r-i), WFC photometry, for galaxies with at least 11 photometric bands and
$\chi^2 > 7$ (red), $<3$ (blue).
R: Histogram of reduced $\chi^2$ for EN1 area, after exclusion of stars.}
\end{figure*}

Figure 12L shows (i-z) versus (r-i) (WFC photometry) for galaxies in EN1 with at least 11 photometric bands,
with different symbols for $\chi^2 < 3$ and $> 7$.  The galaxies with higher $\chi^2$ show a wider spread
in colours, suggesting the main issue is photometric problems.  Figure 12R shows a histogram of the reduced 
$\chi^2$ for sources in the EN1 area, after removal  of stars.

\section{Revised SWIRE Photometric Redshift Catalogue}
Our revised SPRC contains 818555 objects, 204421 in EN1, 116195 in EN2, 217461 in Lockman, and 280478 in XMM-LSS, compared
with 875353 in the same areas in the original SPRC.  3.6$\%$ of SPRC sources did not find a match in the fusion catalogues,
mainly because the latter omitted 24 $\mu$m only sources.  A further 1$\%$ failed to achieve a redshift solution, either because
there were less than two valid photometric bands or because the reduced $\chi^2 >$ 100.  The SWIRE redshifts in CDFS and S1 
have not been revised because we have no new photometric information in these areas.  These two areas bring the total
number of redshifts in the revised catalogue \footnote[4]{ http://astro.ic.ac.uk/$\sim$mrr/swirephotzcat/zcatrev12ff3.dat.gz} to 1009607, with readmeSWIRErev in same directory).  Our new catalogue delivers photometric redshifts for 26288 quasars.

\begin{figure*}
\epsfig{file=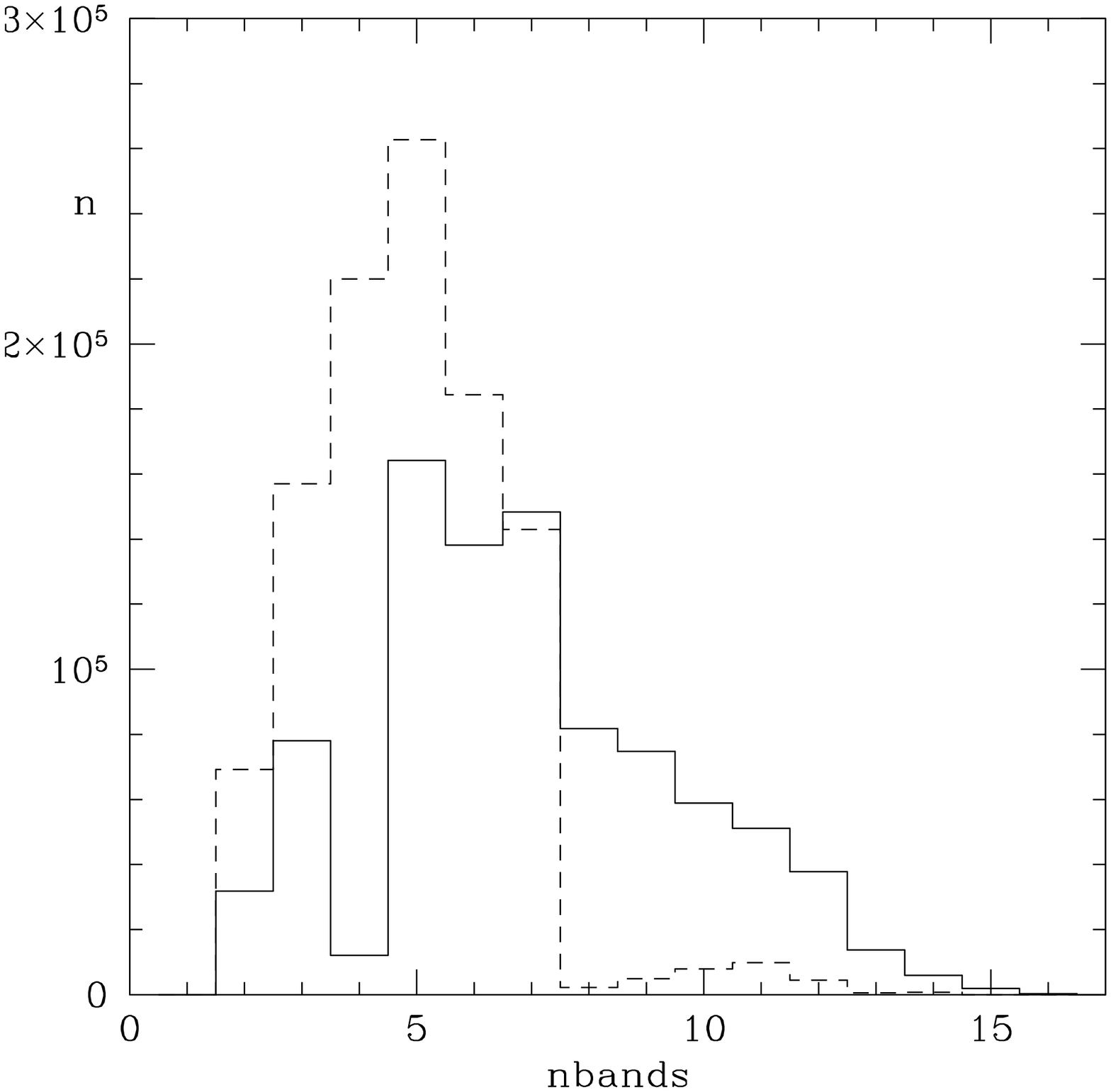,angle=0,width=7cm}
\epsfig{file=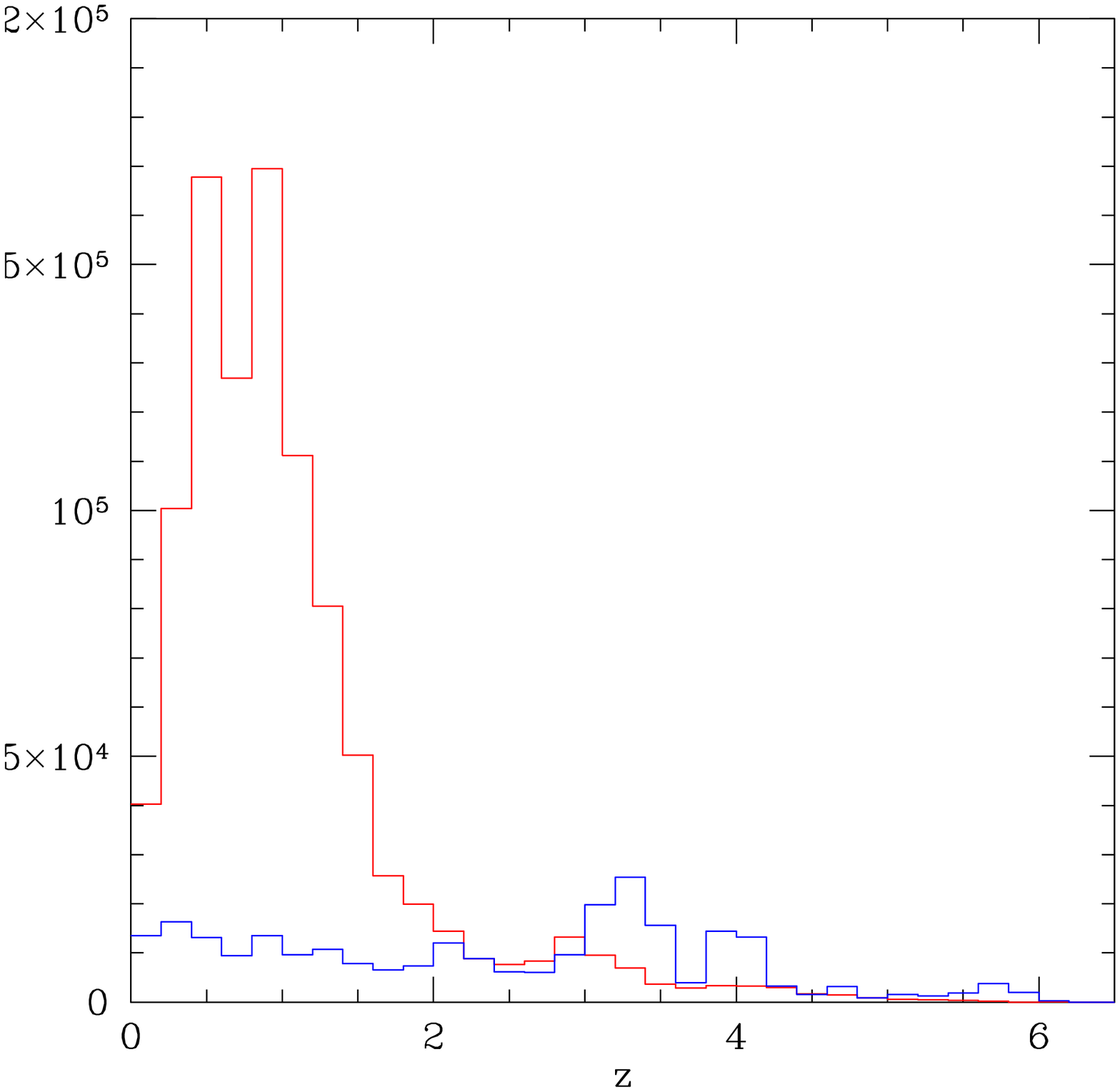,angle=0,width=7cm}
\caption{L: Histogram of number of bands used in estimating photometric redshifts in revised catalogue
(solid line) compared with SPRC (broken line).
R: Histogram of redshifts in revised SWIRE Photometric Redshift Catalogue (red: galaxies, blue: QSOs, x 10).}
\end{figure*}

Fig 13L shows a histogram of the number of photometric bands for the new catalogue, compared with that for
the SPRC.  Fig 13R shows the redshift distribution of galaxies and quasars in the new catalogue.  Although most of the
galaxy redshifts are at z $<$ 1.5, there is a broad tail to higher redshift.  Our catalogue is a very significant resource for galaxies at
high redshift, with 106537 sources with z $>$ 2 and 15226 sources with z $>$ 4.

\begin{figure*}
\epsfig{file=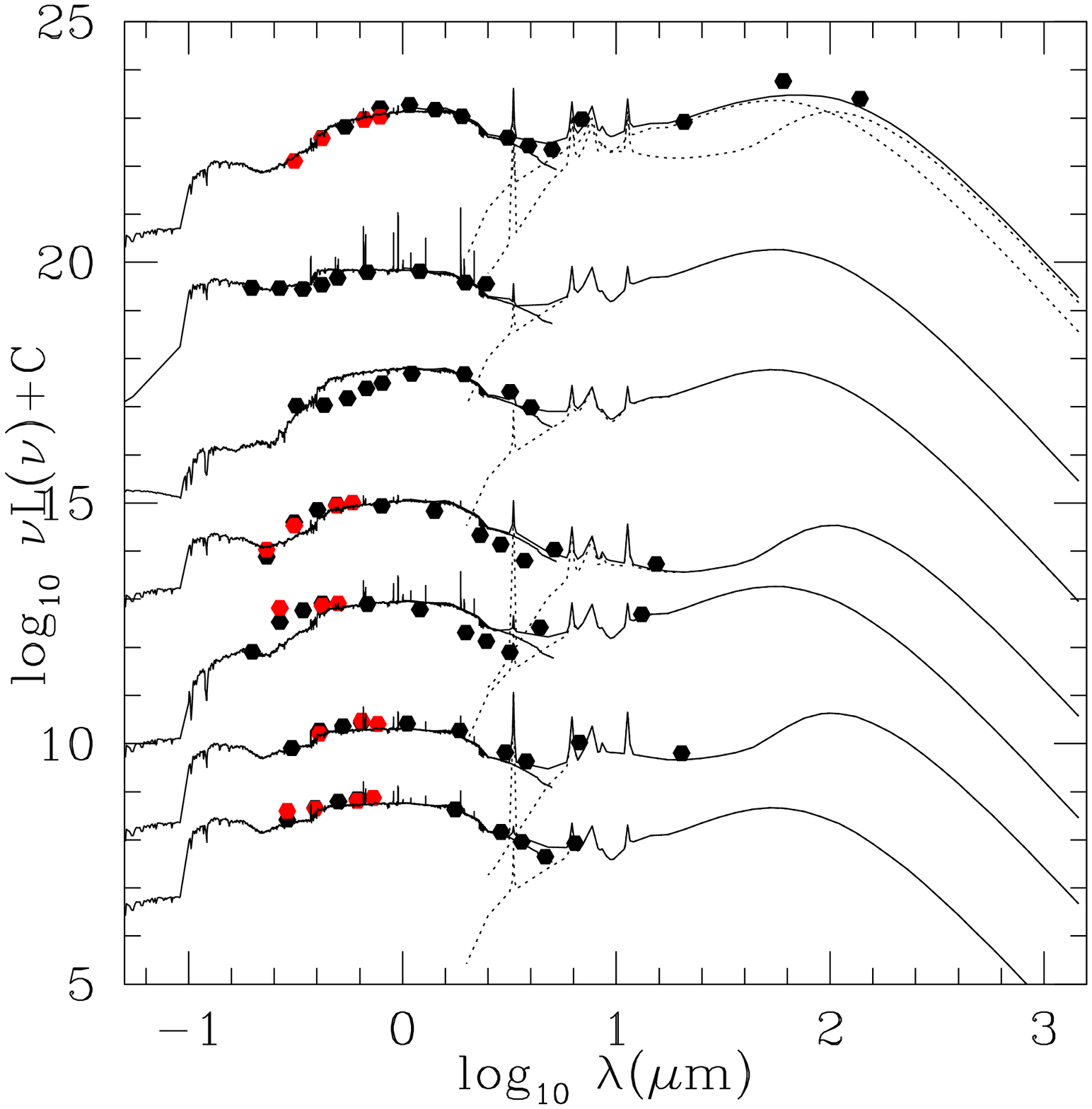,angle=0,width=14cm}
\caption{SEDs of outliers in $z_{ph}$ v. $z_{sp}$ plot, using the spectroscopic redshift, with details of sources given in 
Table 1.  Outliers with more than 12 photometric bands in EN1, EN2, Lockman, and more than 9 bands in XMM-LSS, 
are included.
}
\end{figure*}

\begin{table*}
\caption{$z_{ph}$ v. $z_{sp}$ outliers whose SEDs are plotted in Fig 10, from bottom}
\begin{tabular}{lllllll}
RA & dec & $z_{spec}$ & ref. & $z_{phot}$ & no. of bands & notes\\
 &&&&&&\\
 161.30797  & 58.74822 & 0.2470 &    1 &  0.52 & 12 & alias\\
 161.72533  & 58.99032 & 0.1890  &   1 & 0.55 & 12 & alias\\
 161.74496  & 58.94529 & 0.8220   &  1 &  0.25 & 12 &  zspec error ?\\
 161.94997  & 59.29018 & 0.5600  &   2 &  0.17 & 13 & zspec error ?\\
  36.40352  & -4.39069 & 0.1307   &  3 &  0.85 &  9 & zspec error (single line) ?\\
  36.47719  & -4.36419 & 0.8277  &   3 &  1.24 &  9 & zspec error (single line) ?\\
 248.47900 & 41.48283 & 0.1592  &  4 &  0.47 & 14 & need Sab with Av=0.35\\
 &&&&&&\\
1  Owen et al 2009 &&&&&&\\
2  Rowan-Robinson & et al 2008 &&&&&\\
3  Le Fevre et al 2005 &&&&&&\\
4  Rowan-Robinson & et al 2004 &&&&&\\ 
\end{tabular}
\end{table*}

\section{SEDs of outliers}
To investigate outliers, we have plotted in Fig 14 the SEDs of 8 outliers
from the $z_{phot}$ v. $z_{spec}$ comparisons, those with $\chi^2 <$ 3, and with more than 12 photometric bands
in EN1, EN2 and Lockman, and with more than 9 bands in XMM-LSS.  SDSS photometry (after applying the
aperture correction of section 2.1) is shown in red.  Details of the sources are given in Table 3.
Two objects have aliases at the spectroscopic redshift, one appears to need an Sab template with $A_V$ =0.35 ($A_V$
is set to zero for the Sab template in our code), and 4 could be errors in the spectroscopic redshift (two were based on a single line).

\begin{figure*}
\epsfig{file=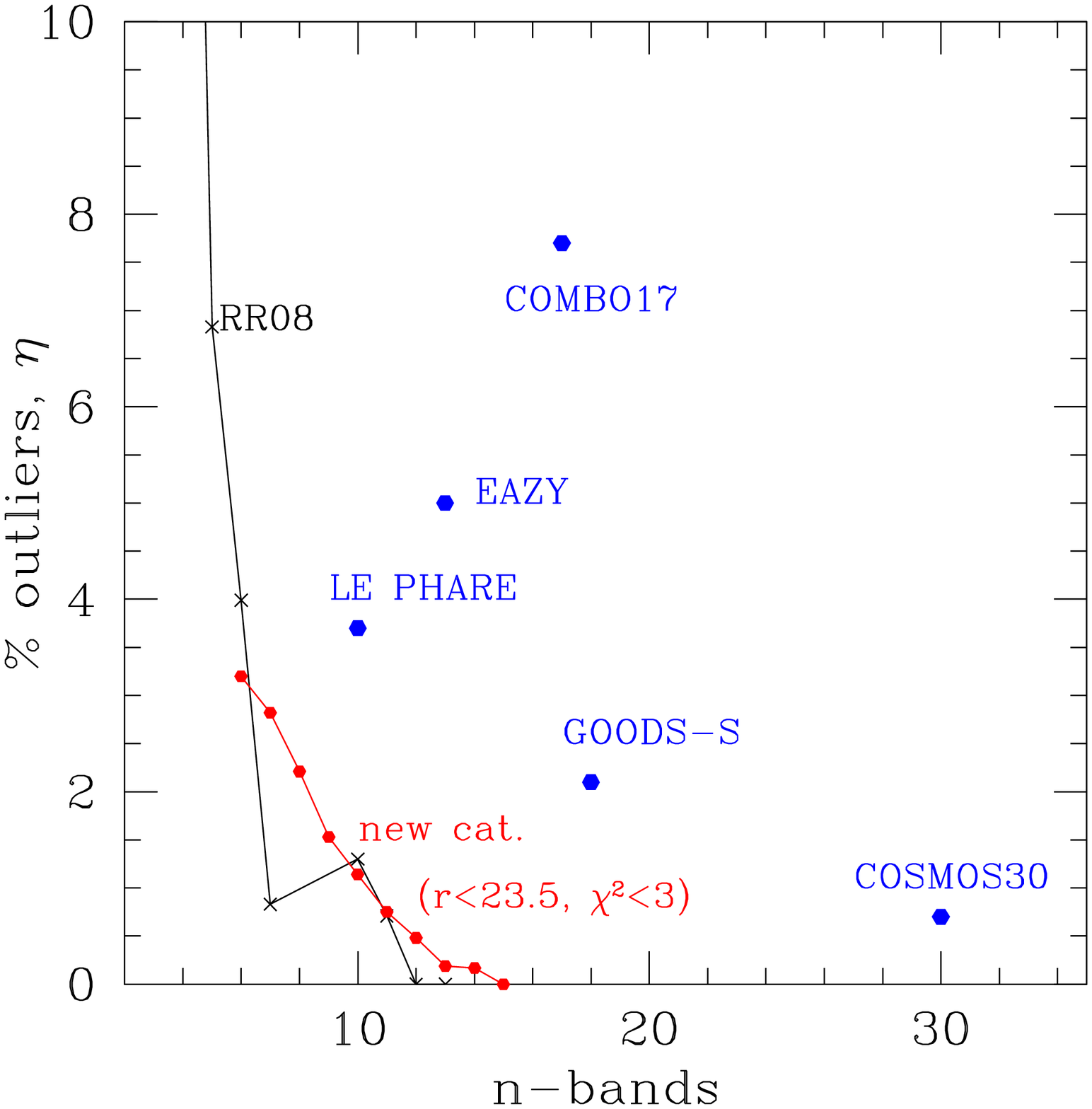,angle=0,width=7cm}
\epsfig{file=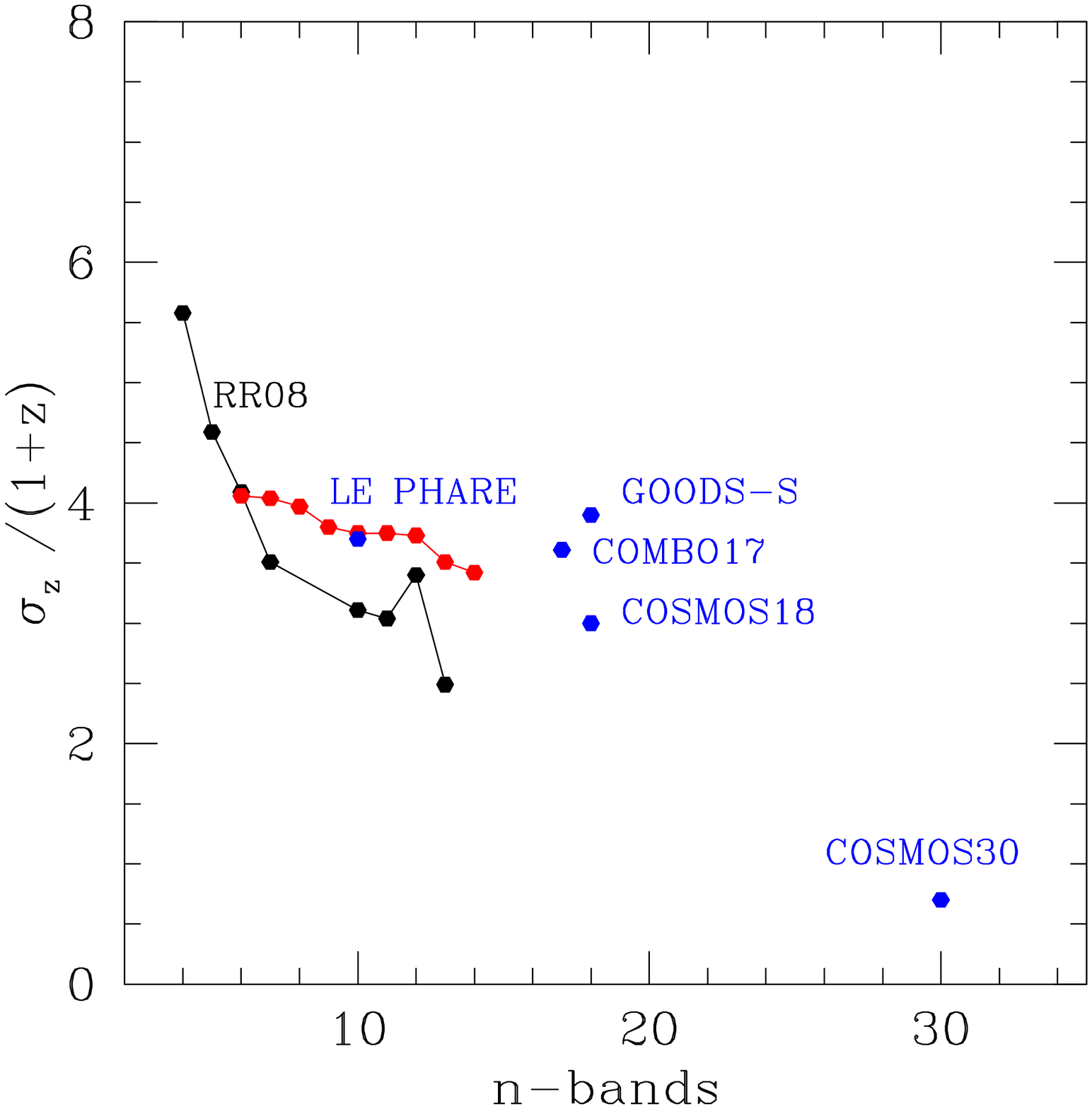,angle=0,width=7cm}
\caption{L: Percentage of outliers versus
number of photometric bands for SPRC (black loci) and for fusion catalogue (red loci), for sources
with reduced $\chi^2 < 3$, $r < 23.5$.
R: Percentage rms ($\sigma_z / (1+z)$) for SPRC (black loci) and for fusion catalogue (red loci), for sources
with reduced $\chi^2 < 3$, $r < 23.5$.}
\end{figure*}

\section{Discussion}

With the changes discussed in the previous sections, we now see improvements in the fraction of catastrophic
outliers in the photometric redshift estimates.  The rms errors for galaxies are very similar to those achieved in
SPRC, whole those for quasars are significantly improved.  

For galaxies with reduced $\chi^2 < 3, r < 23.5$, the rms in $(z_{phot}-z_{spec})/(1+z_{spec})$, after 
rejection of outliers with values discrepant by 15$\%$ or more, is
3,7, 3.4 $\%$, for no. of bands 10, 14 respectively.  The corresponding 
percentages of outliers are 1.2 and 0.2 $\%$, respectively.  
Fig 15L shows how the percentage of outliers for galaxies with
$|log_{10} (1+z_{phot})/ (1+z_{spect})| > 0.06$, vary with the number of photometric bands.  Fig 15R shows the
percentage rms ($\sigma_z / (1+z)$)  versus number of photometric bands.  
The additional photometric bands provided by the JHK data from 2MASS and UKIDSS now have a clear
beneficial effect, especially on the fraction of catastrophic outliers.  Our results can be 
compared directly with the corresponding results for SPRC, and with new results from EAZY (Brammer et al 2008), GOODS-S (Dahlen et al 2010),
and COSMOS30 (Ilbert et al 2009).  While these comparisons can be affected by differences in the survey depths, the main differences
between these surveys is the number of photometric bands.  Our outlier performance is consistently better than other methods.

Our new SWIRE redshift catalogue should be useful for improved studies of the infrared extragalactic population and for
the Herschel surveys carried out in all of the SWIRE fields by the Hermes consortium.


\end{document}